\documentclass{aa}

\usepackage{graphicx}
\usepackage{txfonts}
\usepackage{subcaption}
\usepackage{lscape}
\usepackage{placeins}
\usepackage{CJKutf8}
\usepackage{color}

\usepackage{cuted}
\usepackage{capt-of}
\usepackage{hyperref}

\newcommand{\fg}[1]{Fig.~\ref{fig:#1}}

\newcommand{\tb}[1]{Table~\ref{tab:#1}\xspace}

\newcommand{\se}[1]{Sect.~\ref{sec:#1}\xspace}
\newcommand{\Se}[1]{Section~\ref{sec:#1}\xspace}

\newcommand{\App}[1]{Appendix~\ref{app:#1}\xspace}

\begin{document}

   \title{Azimuthal molecular variations in the AB~Aur planet-forming disk}

   \subtitle{}

   \author{
   Haochang Jiang \begin{CJK*}{UTF8}{gbsn}(蒋昊昌)\end{CJK*}\inst{1}
   \and
   Dmitry Semenov\inst{2,1}
   \and
   Myriam Benisty\inst{1}
   \and
   Vincent Pi\'etu\inst{3}
   \and
   Thomas Henning\inst{1}
   \and
   Pablo Rivi\`ere-Marichalar \inst{4}
   \and
   Lucas M. Stapper\inst{1}
   \and
   Edwige Chapillon\inst{3}
   }

   \institute{
   Max-Planck Institute for Astronomy, Königstuhl 17, 69117 Heidelberg, Germany\\
   \email{h-jiang@mpia.de}
   \and
   Institut für Theoretische Astrophysik, Zentrum für Astronomie der Universität Heidelberg, Albert-Ueberle-Str. 2, 69120 Heidelberg, Germany
   \and 
   Institut de Radioastronomie Millimétrique (IRAM), 300 Rue de la Piscine, F-38046 Saint Martin d'H\`{e}res, France
   \and Centro de Astrobiología (CAB), CSIC-INTA, Ctra Ajalvir Km 4, Torrejón de Ardoz, 28850 Madrid, Spain
   }

  \abstract
  {
  Late infall episodes, where material from the surrounding environment accretes onto Class~II protoplanetary disks, are emerging as a potentially important but poorly quantified driver of disk evolution. Observed as filamentary streamers in molecular lines and scattered light, such late-stage accretion can perturb disk structures through localized shocks, density enhancements, and warps, yet its chemical consequences remain poorly constrained. We present NOEMA 1.2\,mm line-survey observations of the AB~Aurigae system, a structured, young Class~II Herbig disk that shows evidence for both ongoing infall and planet formation. We detect strong azimuthal chemical diversity: SO emission is enhanced in the northern disk near the inferred streamer--disk interaction region, while C$_2$H emission peaks on the opposite southern side; in contrast, CS forms a nearly axisymmetric ring. HCN and HCO$^+$ instead peak near the dust continuum overdensity at the main mm-sized dust ring. Using multi-transition rotational-diagram analyses of SO and CS, we quantify the azimuthal contrast in column density and excitation. The SO-bright sector exhibits higher rotational temperatures and SO column densities, whereas CS remains nearly axisymmetric with substantially lower rotational temperatures, suggesting that the two species probe different disk layers and/or chemical components. For C$_2$H, potential temperature variations contribute to, but cannot fully explain the observed asymmetries. The HCO$^+$/H$^{13}$CO$^+$ line ratio further indicates that HCO$^+$ is optically thick across the molecular ring, while the elevated ratio inside the cavity suggests an enhanced gas-phase $^{12}$C/$^{13}$C, consistent with isotope-selective photodissociation. Comparison with gas-grain chemical models favors gas-phase C/O ratios near or above unity, with higher effective C/O in the C$_2$H-bright sector. We discuss two plausible, non-exclusive origins for the observed chemical asymmetries: (i) infall-induced heating and desorption of O-bearing ices that enhance SO and lower the local gas-phase C/O near the streamer's impact site, and (ii) planet-driven substructures and localized heating or enhanced UV irradiation that can promote hydrocarbon-rich chemistry on the opposite side. These results highlight that environmental accretion and planet formation can jointly imprint azimuthal variations in disk chemistry, with potential consequences for the compositions of forming planets.
  }


   \maketitle
   \nolinenumbers

\section{Introduction}\label{sec:introduction}

Protoplanetary disks inherit materials from their natal environments while simultaneously reprocessing them through both physical and chemical evolution. Grain growth, radial drift, and planet formation redistribute solids throughout the disk \citep[e.g.,][]{Weidenschilling1977a,BrauerEtal2008,PinedaEtal2023,DrazkowskaEtal2023,Birnstiel2024}, while gas-phase reactions, ice chemistry, freeze-out, desorption, and irradiation continuously modify the molecular composition \citep[e.g.,][]{EistrupEtal2018}. Together, these processes redistributes carbon- and oxygen-bearing volatiles between gas and solids, thereby modifying the gas-phase elemental carbon-to-oxygen ratio (C/O). Since C/O regulates the abundances of many simple molecules and radicals, it provides a powerful diagnostic of disk chemistry and may ultimately influence the bulk composition of forming planetary atmospheres \citep[e.g.,][]{OebergEtal2011,BerginEtal2024b}.

A radial C/O gradient in the gas phase is a natural outcome of snowline physics and solid--gas (de)coupling, and is now routinely predicted by disk chemical models. In contrast, azimuthal chemical variations have been observed in only a few cases, whose origin and implications are poorly understood. Because such asymmetries require departures from axisymmetry, they are generally interpreted as signatures of localized dynamical processes, such as late infall, spiral shocks, dust traps, and embedded planets. Most previous studies have focused on disks with extreme dust asymmetries, such as IRS~48 and HD~142527, where chemical contrasts are associated with prominent dust traps \citep{BoothEtal2021b,TemminkEtal2023}. Whether more moderately asymmetric but dynamically active disks exhibit comparable azimuthal chemical structure is still unclear.

The Herbig~Ae star AB~Aurigae \citep[$M_\star=2.23\,M_\odot$, $\tau=3.9$--4.4~Myr, $d=156$~pc;][]{SpeedieEtal2024,GarufiEtal2024,GaiaCollaborationEtal2021} provides an ideal laboratory to investigate this question. The disk hosts large-scale spirals and kinematic deviations from Keplerian rotation \citep{TangEtal2017,BoccalettiEtal2020}, and shows evidence for filamentary streamers intersecting the outer disk \citep{DutreyEtal2024,SpeedieEtal2025}. At the same time, signatures of ongoing planet formation have been reported in both gas and scattered-light observations \citep[e.g.,][]{CurrieEtal2022,CurrieEtal2025}. The coexistence of environmental accretion and internal disk substructures makes AB~Aur uniquely suited to test how external and internal processes jointly shape disk chemistry.

Several molecular studies have been conducted toward AB Aur. Using IRAM 30~m telescope, \citet{SemenovEtal2005} conducted the first line survey on AB~Aur and detected CS, HCO$^+$, DCO$^+$, H$_2$CO, HCN, HNC, and SiO around the disk. Also with IRAM 30~m antenna, \citet{FuenteEtal2010} detected additionally CN and SO, and tentatively detected H$_2$S, marking the first detection of SO in a protoplanetary disk. More detailed studies of SO as shock tracer in AB Aur have been conducted by \citet{DutreyEtal2024} using high angular resolution ALMA data, and further demonstrated by \citet{SpeedieEtal2025} to be related with the infall. \citet{FuenteEtal2010} also determined the upper limit of a number of molecules in AB~Aur, including C$_2$H, which is detected by our new Northern Extended Millimetre Array (NOEMA) observations and will be the focus of this work. Meanwhile, a long-term NOEMA line survey on it has returned a fruitful set of resolved molecular mapping, including the studies of gas accretion flow through HCN and HCO$^+$, the sulfur budget, and C/O ratio based on CS and SO \citep{Rivi`ere-MarichalarEtal2019,Riviere-MarichalarEtal2020,Riviere-MarichalarEtal2022,Rivi`ere-MarichalarEtal2026}.

\begin{figure*}
    \centering
    \includegraphics[height=.99\textheight]{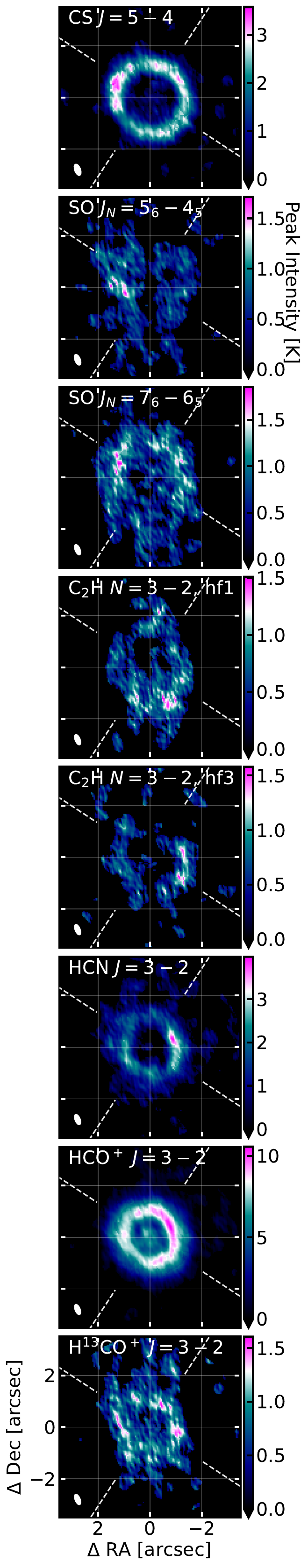}
    \includegraphics[height=.99\textheight]{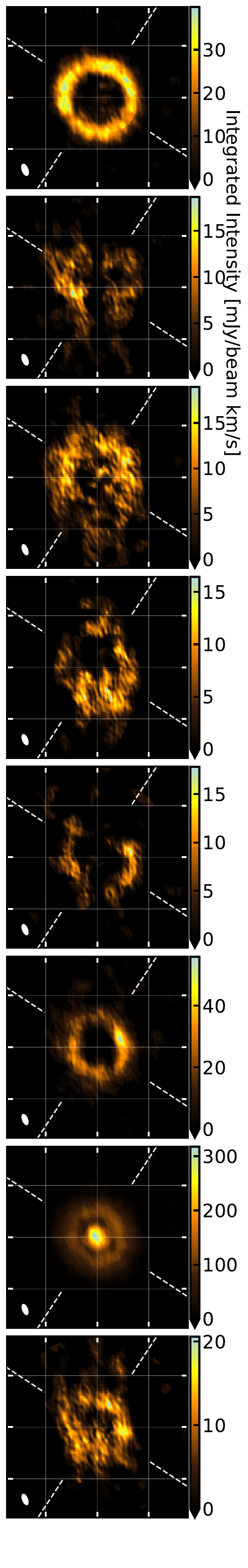}
    \includegraphics[height=.99\textheight]{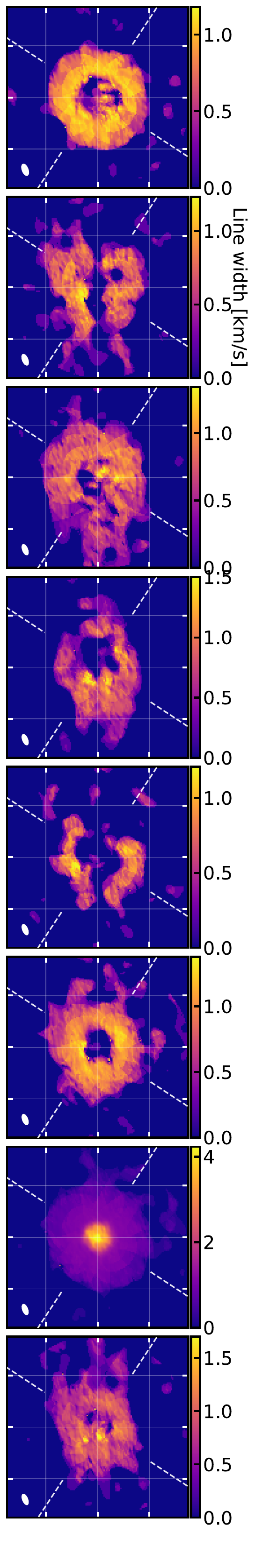}
    \includegraphics[height=.99\textheight]{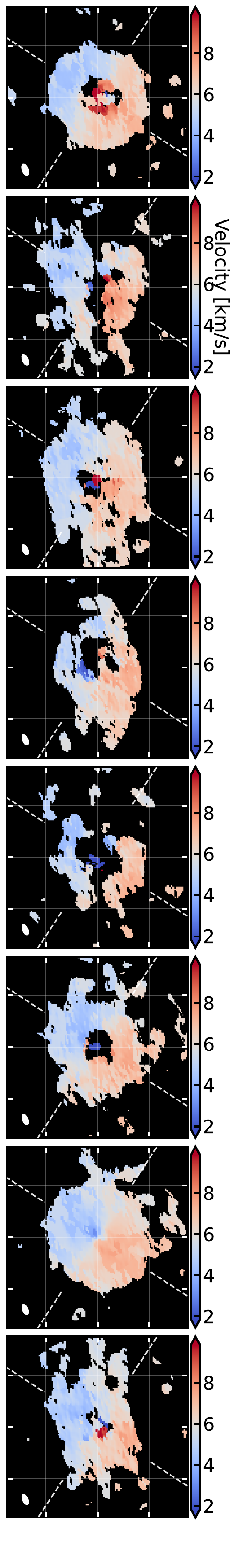}
    \caption{
    The gallery of detected lines in peak intensity, integrated intensity, line width and line central velocity maps (from left to right).
    }
    \label{fig:AB_Aur_maps}
\end{figure*}

\begin{figure}
    \centering
    \includegraphics[height=.88\textheight]{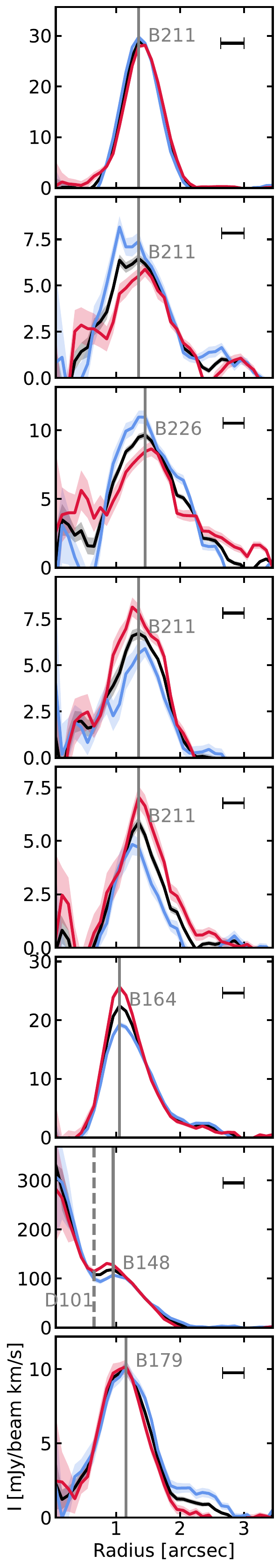}
    \includegraphics[height=.88\textheight]{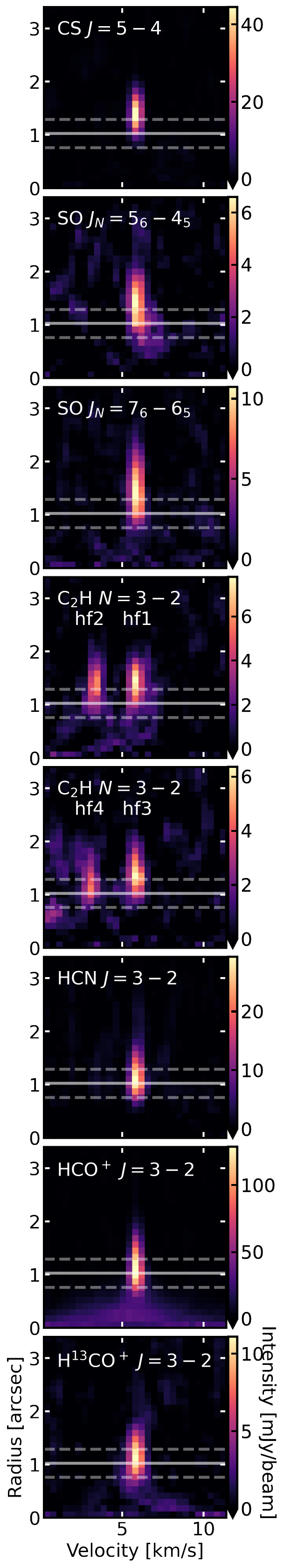}
    \caption{
    {\bf Left}: De-projected, azimuthally averaged radial profiles of the full-disk (black), and along the red- and blue-shift sides of the disk (red and blue). In each panel, the radial locations of the ring peak and the gap bottom are labeled "BX" and "DX," respectively, where X is the distance in au.
    {\bf Right}: Teardrop plots. The solid line indicates the radial location of the continuum ring. And the dashed line marks the FWHM of the continuum ring.  
    }
    \label{fig:AB_Aur_profiles}
\end{figure}

In this work, we present high angular-resolution NOEMA observations targeting SO, C$_2$H, CS, HCN, HCO$^+$, and H$^{13}$CO$^+$ at 1.2~mm (\Se{observation}). To our knowledge, these are the first detections of C$_2$H and H$^{13}$CO$^+$ in AB~Aur. We report a pronounced azimuthal anti-correlation between SO and C$_2$H: SO peaks in the north of the disk near the inferred streamer--disk interaction region \citep{SpeedieEtal2025}, while C$_2$H peaks on the opposite side. CS~$J{=}5{-}4$ forms a nearly axisymmetric ring at radii comparable to SO and C$_2$H, whereas HCN, HCO$^+$ and H$^{13}$CO$^+$ are brightest near the continuum overdensity along the main dust ring. This multi-species comparison enables us to disentangle chemical effects associated with large-scale infall from those linked to local dust substructures.

To interpret these spatial variations, we analyze the moment maps and azimuthally resolved radial profiles of the line emission (\Se{asymmetry}) and derive apparent column densities under the assumption of Local Thermal Equilibrium (LTE). We then compare the observational constraints to a grid of gas-grain chemical models with varying elemental C/O ratios in \se{model}. In \Se{discussions}, we assess two physically motivated scenarios: (i) infall-induced heating and desorption of O-bearing ices that enhance SO near the streamer impact site, and (ii) planet-driven substructures and localized heating that promote hydrocarbon-rich chemistry on the opposite side of the disk. Using this combined observational and modeling approach, we explore whether AB~Aur hosts a chemically heterogeneous disk and identify the physical processes responsible. We conclude by outlining testable predictions and implications for the atmospheric compositions of wide-separation companions forming in chemically heterogeneous disks. The major findings are summarized in \se{summary}.

\section{NOEMA observations} \label{sec:observation}

AB~Aur was observed with NOEMA under programme W24BR (PI: H.~Jiang) using the PolyFiX correlator in dual-polarization mode. The frequency setup covered 244.2–252.3\,GHz and 259.7–267.8\,GHz, providing continuous 8+8\,GHz bandwidth. The wide bandwidth enables coverage of numerous molecular transitions; in this work, we focus on the species with robust detections, namely CS, SO, C$_2$H, HCN, HCO$^+$, and H$^{13}$CO$^+$. Molecular data for all targeted transitions are listed in \tb{transition}.

The observations were conducted in line-survey mode with a uniform spectral resolution of 250\,kHz across the full bandwidth, corresponding to $\sim0.3$\,km\,s$^{-1}$ at 250\,GHz. Spectral windows containing the targeted lines were extracted and regridded to a uniform channel spacing of 0.4\,km\,s$^{-1}$ for imaging.

Observations were carried out in three tracks: 2.4\,h on source on 27 December 2024 in the C configuration (12 antennas; baselines 24–368\,m), and 2.4\,h and 0.9\,h on source on 15 and 17 February 2025 in the extended A configuration (12 antennas; baselines 72–1664\,m). The bandpass was calibrated using 3C84. J0418$+$380 and J0438$+$300 were used as phase calibrators. Absolute flux calibration was obtained from LkH$\alpha$~101 (first and third tracks) and MWC~349 together with 2010$+$723 (second track). The absolute flux uncertainty is $\sim10\%$\footnote{NOEMA technical reports, \href{https://www.mn.uio.no/astro/english/services/it/help/astronomy-software/gildas/astro-planet-models.pdf}{\textit{Planet brightness temperatures in ASTRO}} V1.0 (2024), J. Boissier, J. Pety, S. Bardeau}.

Calibration was performed using the standard NOEMA pipeline within \texttt{GILDAS} \citep{GildasTeam2013}. After inspection and flagging of low-quality data, continuum $uv$ tables were produced by identifying and excluding channels with strong line emission, and averaging the remaining line-free channels within each sideband. Three rounds of phase-only self-calibration were applied to the continuum data with \texttt{IMAGER}. The final gain solutions were transferred to the line spectral windows within the same basebands. Continuum subtraction was performed in the $uv$-plane using a linear baseline fit.

The calibrated data were exported to measurement sets and imaged in CASA 6.6.4 \citep{CASATeamEtal2022}. Line cubes were generated with a channel spacing of 0.4\,km\,s$^{-1}$ using the \texttt{multiscale} deconvolver with scales of [$0''$, $0\farcs25$, $0\farcs75$, $1\farcs75$, $5\farcs0$]. We adopted the same iterative masking procedure used in exoALMA data production \citep{LoomisEtal2025}. After testing different weighting schemes, we selected a Briggs robust parameter of 1.5 to balance angular resolution and sensitivity. Primary-beam correction was applied to all final images, though given the size of the disk, the impact is minor. The resulting synthesized beams range from $0\farcs46\times0\farcs23$ to $0\farcs50\times0\farcs26$ (\tb{transition}). The rms noise per 0.4\,km\,s$^{-1}$ channel is 2.9–4.7\,mJy\,beam$^{-1}$, depending on frequency.

Peak intensity, integrated intensity (moment~0), line width and peak velocity maps were constructed using a Keplerian mask (see \App{channel_maps}) extending to $3\farcs5$, generated with \texttt{bettermoments} \citep{TeagueForeman-Mackey2018} and presented in \fg{AB_Aur_maps}. The mask includes emission detected above $1\,\sigma_{\rm map}$ in at least two consecutive channels, where $\sigma_{\rm map}$ is measured from nearby line-free regions. This approach minimizes noise bias while retaining extended emission. The peak intensity and velocity maps are generated using the \texttt{quadratic} method presented in \citet{TeagueForeman-Mackey2018}. And the line width maps are calculated with the \texttt{width} method of \texttt{bettermoments}, which gives the effective line width $\Delta V = M_0/(\sqrt{\pi}F_\nu)$, where $M_0$ is the integrated intensity and $F_\nu$ is the peak intensity. These methods are more robust against noise than the traditional moment 8, 9, and 2 maps.

Line fluxes were measured from spectra extracted within the Keplerian mask and integrated over the velocity range 0.2-11.8\,km\,s$^{-1}$ for HCO$^+$ and 4.2–7.4\,km\,s$^{-1}$ for all other lines, see \App{channel_maps}. Flux uncertainties were estimated as $\sqrt{N}\,\Delta v\,\sigma_{\rm spec}$, where $N$ is the number of integrated channels, $\Delta v$ the channel width, and $\sigma_{\rm spec}$ the rms of line-free channels. The measured integrated fluxes are listed in \tb{transition}. We also use the same procedure for archival data used in our analysis. The corresponding data information is listed in \tb{transition_archive},  and see \App{archival_data}.

All detected molecular lines show ring-like emission morphologies at radii of $\sim150$–220\,au. CS~$J{=}5{-}4$ forms a nearly axisymmetric ring. SO and C$_2$H display pronounced azimuthal asymmetries, with their emission peaks located on opposite sides of the disk. HCN and HCO$^+$ peak near the dust continuum overdensity along the main ring and, in the case of HCO$^+$, additionally reveal spatially resolved emission inside the dust cavity \citep[see also][]{Rivi`ere-MarichalarEtal2019}. Radial profiles of line emission are shown in \fg{AB_Aur_profiles}, where `B' denotes the brightness peak of the ring and `D' the gap minimum in unit of au. The alignment with the continuum ring is illustrated in the teardrop plots (right panels of \fg{AB_Aur_profiles}), where the white horizontal line marks the continuum peak radius and the dashed lines indicate its FWHM.

\begin{table*}
\centering
\caption{Molecular data.}
\label{tab:transition}
\makebox[\textwidth][c]{
\begin{tabular}{l l l l c c c c c c c c}
\hline\hline
Molecule & Transition & Rest freq. & $\log_{10}(A_{ij}/{\rm s^{-1}})$ & $E_u$ & $g_u$ & Beam size (PA) & rms noise & Integrated flux \\
 & & [GHz] & & [K] & & [$''\times''$] ([deg]) & [mJy bm$^{-1}$] &  [mJy km/s] \\

(1) & (2) & (3) & (4) & (5) & (6) & (7) & (8) & (9)\\
\hline
CS            & $J=5-4$                                      & 244.93556  & $-3.52707$ & 35.3 & 11 & 0.50$\times$0.26 (21) & 3.0 & 1168$\pm$39
 \\
SO            & $J_N=5_6-4_5$                                & 251.82577    & $-3.71555$ & 50.7 & 11 & 0.48$\times$0.25 (22) & 2.9 & 492$\pm$54
 \\
SO            & $J_N=7_6-6_5$                                & 261.84372   & $-3.64166$ & 47.6 & 15 & 0.47$\times$0.23 (22) & 3.0 & 887$\pm$79
 \\
C$_2$H$^a$    & $N=3-2,\ J=\tfrac{7}{2}-\tfrac{5}{2},\ F=4-3$        & 262.00426    & $-4.27528$ & 25.1 & 9  & 0.47$\times$0.23 (22) & 3.0 & 427$\pm$71
 \\
C$_2$H        & $N=3-2,\ J=\tfrac{7}{2}-\tfrac{5}{2},\ F=3-2$        & 262.00648   & $-4.29203$ & 25.1 & 7  & 0.47$\times$0.23 (22) & 3.0 & 295$\pm$57
 \\
C$_2$H        & $N=3-2,\ J=\tfrac{5}{2}-\tfrac{3}{2},\ F=3-2$        & 262.06499   & $-4.31202$ & 25.2 & 7  &  0.47$\times$0.23 (22) & 3.0 & 310$\pm$57
 \\
C$_2$H        & $N=3-2,\ J=\tfrac{5}{2}-\tfrac{3}{2},\ F=2-1$        & 262.06747   & $-4.35069$ & 25.2 & 5  & 0.47$\times$0.23 (22) & 3.0 & 191$\pm$57
 \\
HCN           & $J=3-2$                                      & 265.88643  & $-3.07662$ & 25.5 & 21 & 0.46$\times$0.24 (22) & 3.6 & 1196$\pm$56
\\
HCO$^+$       & $J=3-2$                                      & 267.55763  & $-2.83764$ & 25.7 & 7  & 0.46$\times$0.23 (21) & 4.7 & 6698$\pm$381
\\
H$^{13}$CO$^+$ & $J=3-2$                                     & 260.25534   & $-2.87375$ & 25.0 & 7  & 0.47$\times$0.23 (22) & 2.9 & 590$\pm$64
\\
\hline
\end{tabular}
}
\tablefoot{
Columns: 
(1) Molecular species; 
(2) Quantum numbers of the observed transition; 
(3) Rest frequency in GHz; 
(4) $\log_{10}(A_{ij}/{\rm s^{-1}})$; 
(5) Upper-state energy $E_u$ in K; 
(6) Upper-state degeneracy $g_u$; 
(7) Synthesized beam size (major $\times$ minor axis in arcsec) and position angle in degrees; 
(8) rms noise per 0.4\,km\,s$^{-1}$ channel in mJy\,beam$^{-1}$; 
(9) Integrated line flux measured within the Keplerian mask over 4.3–7.3\,km\,s$^{-1}$ in mJy\,km\,s$^{-1}$. 
Molecular spectroscopic parameters are taken from the Cologne Database for Molecular Spectroscopy (CDMS) \citep{MuellerEtal2001}. \\
$^{a}$ For simplicity, the four C$_2$H $N=3{-}2$ hyperfine components are hereafter referred to as hf1, hf2, hf3, and hf4, in order of increasing rest frequency and decreasing Einstein A coefficients.
}
\end{table*}

\section{Results}\label{sec:asymmetry}

\subsection{Morphology of imaging}

The moment~0 and peak intensity maps reveal pronounced azimuthal contrasts in SO and C$_2$H. Both newly observed SO transitions ($J_N=5_6-4_5$ and $7_6-6_5$) and the previously reported $6_5-5_4$ line exhibit enhanced emission in the northern sector of the disk \citep{DutreyEtal2024}. This enhancement spatially coincides with the streamer--disk interaction region identified in CO and scattered light observations \citep{SpeedieEtal2025}. The consistent morphology across multiple SO transitions indicates that the asymmetry is not confined to a single upper-state energy.

In contrast, the four detected C$_2$H $N=3-2$ hyperfine components peak on the opposite (southern) side of the disk. To improve signal-to-noise, we stacked the visibilities of the four hyperfine components (hf1–hf4; \tb{transition}) following \citet{HuangEtal2024}. The stacked image confirms a clear anti-correlation between SO and C$_2$H (\fg{SO_C2H}).
The azimuthal contrast is also evident when integrating radial profiles separately along the blue- and red-shifted sides of the disk (left panels of \fg{AB_Aur_profiles}).

\begin{figure*}
    \centering
    \begin{minipage}[c]{0.25\linewidth}
        \centering
        \includegraphics[width=\linewidth]{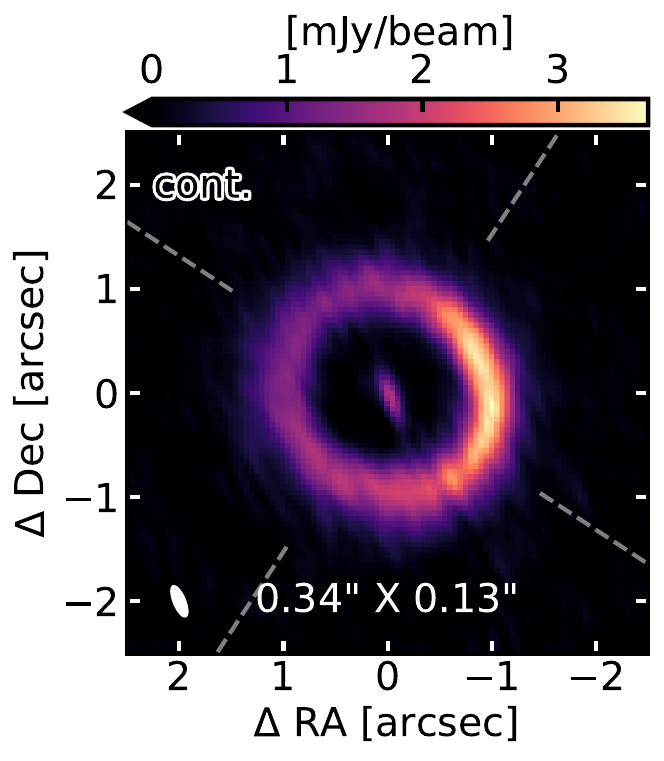}
    \end{minipage}%
    \hfill
    \begin{minipage}[c]{0.75\linewidth}
        \centering
        \includegraphics[width=0.32\linewidth]{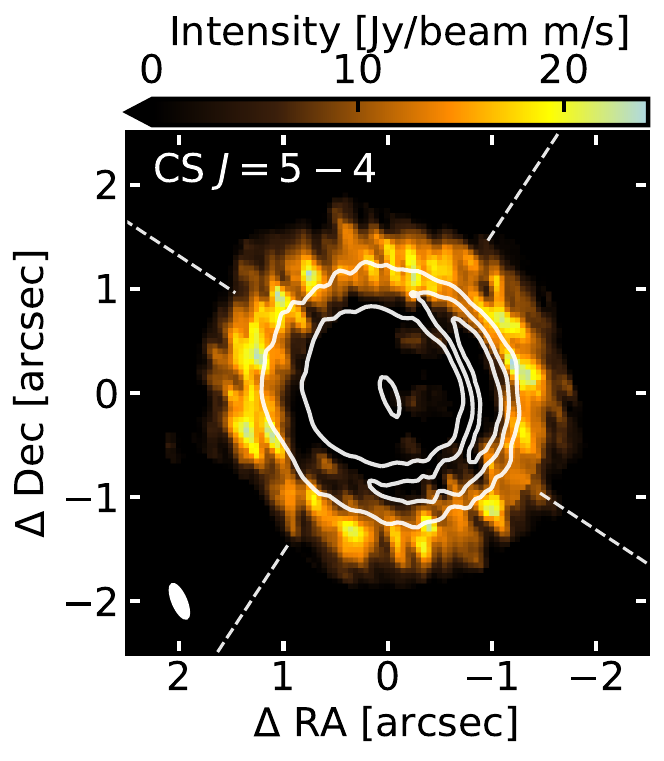}
        \includegraphics[width=0.32\linewidth]{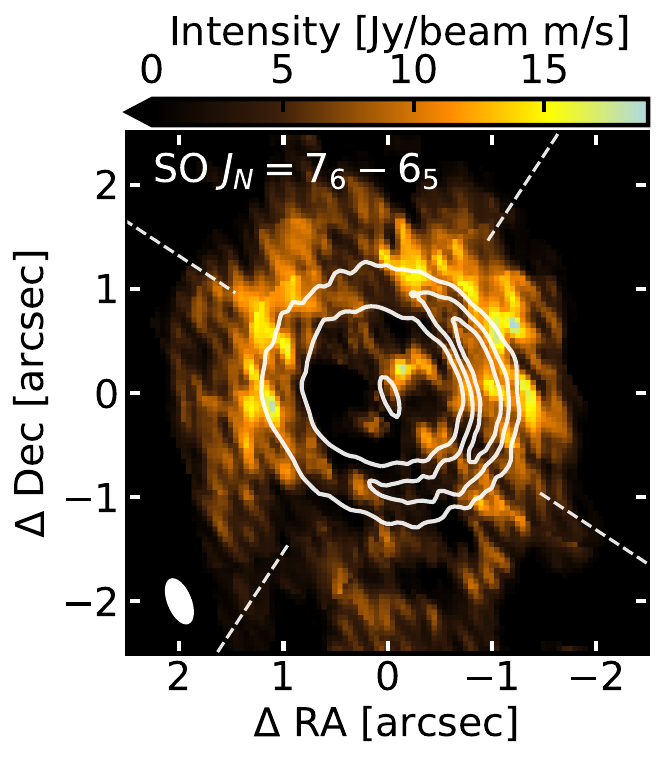}
        \includegraphics[width=0.32\linewidth]{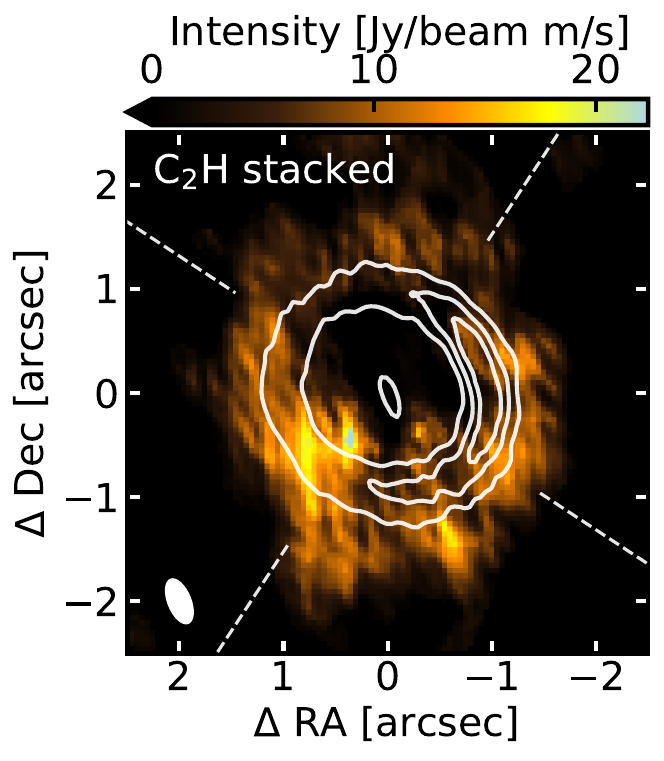}
        \includegraphics[width=0.32\linewidth]{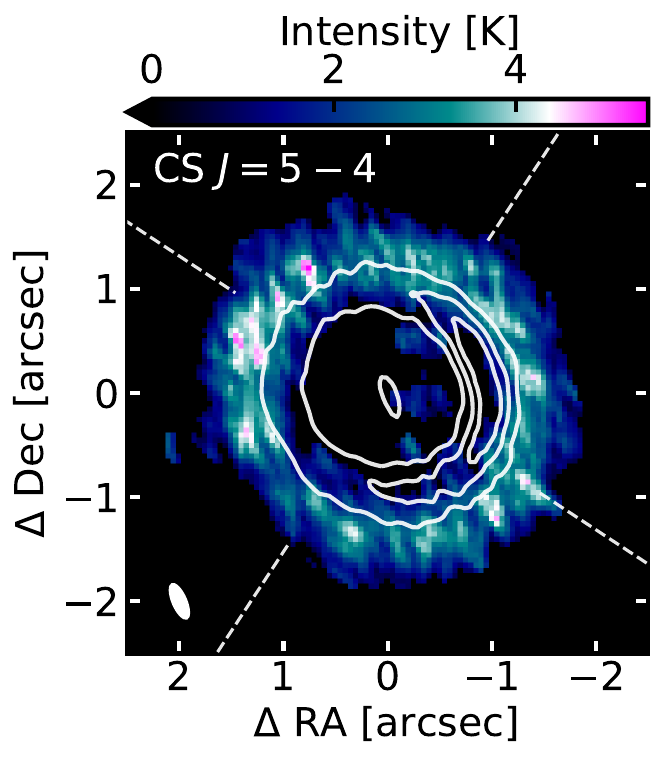}
        \includegraphics[width=0.32\linewidth]{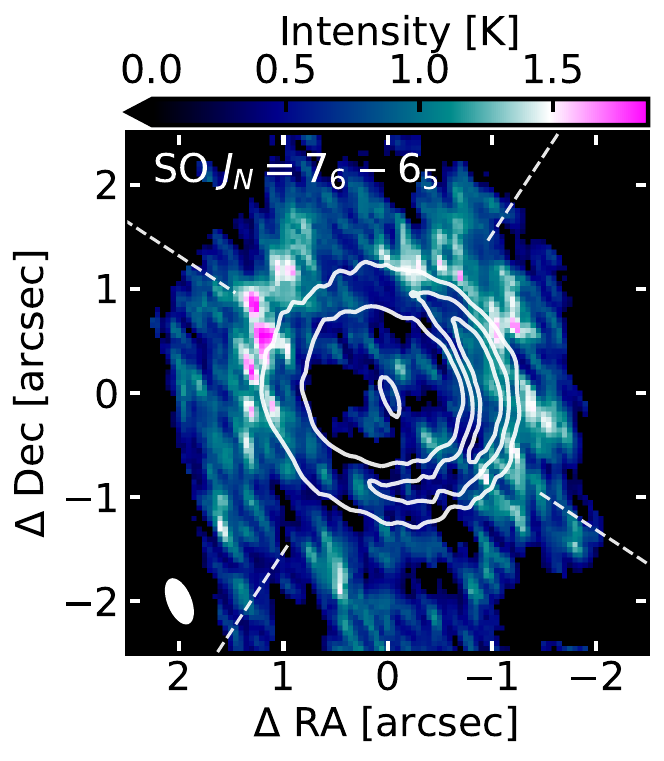}
        \includegraphics[width=0.32\linewidth]{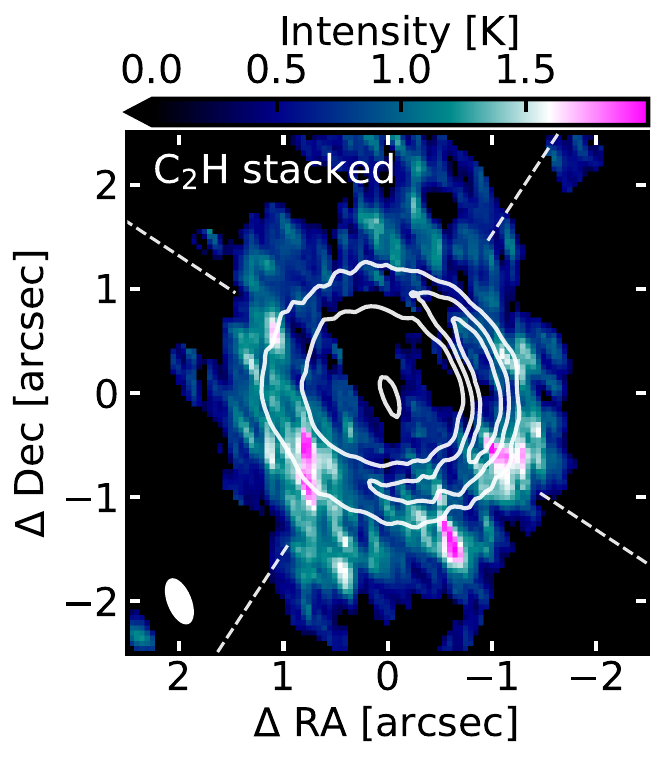}
    \end{minipage}
    \caption{Integrated intensity ({\bf Top}) and peak intensity ({\bf Bottom}) maps of CS $J=5-4$, SO $J_{N} = 6_{5}-5_{4}$ and stacked C$_2$H. Overlaid contours show the r $=0.5$ 266\,GHz continuum ({\bf Left}, see also \App{continuum}) at 15, 30, and 45$\sigma$, where $\sigma=0.06$\,mJy\,beam$^{-1}$ is the rms noise measured at emission free region. The CS maps shown are imaged with r $=0.5$ for better angular resolution. Dashed lines indicate the disk major and minor axes. }
    \label{fig:SO_C2H}
\end{figure*}

CS~$J{=}5{-}4$, which traces similar radii ($\sim150$–220\,au), forms a nearly axisymmetric ring in integrated map. Though a small enhancement in appears in the quadratic fit to peak intensity map in the northeast (\fg{AB_Aur_maps}), this difference is not significant in the momentum zero map nor the radial profile between the blue- and red-shifted regions (\fg{AB_Aur_profiles}). The absence of a strong CS asymmetry is notable given the pronounced contrasts in SO and C$_2$H. 

HCN and HCO$^+$ show yet another morphology: both peak near the dust continuum overdensity along the main ring (\fg{HCN_HCOp}), suggesting that their azimuthal variation is more closely linked to local dust substructure than to the large-scale streamer interaction. The behavior of HCN and HCO$^+$ in AB~Aur is thus very similar to those of IRS~48 \citep{BoothEtal2024b} and HD~142527 \citep{TemminkEtal2023}, which might be related to the potentially icy dust trap. In addition, HCO$^+$ is the only specie where the inner gas disk is unambiguously detected. Its existence was first reported with low resolution in \citet{Rivi`ere-MarichalarEtal2019}. The HCO$^+$ peak is slightly offset from the central continuum emission toward the blue-shifted northeastern side, which is likely due to intrinsic asymmetry between the blue- and red-shifted parts of the inner disk, see \App{hcop_inner_disk}. 

\begin{figure*}
    \centering
    \begin{minipage}[c]{0.25\linewidth}
        \centering
        \includegraphics[width=\linewidth]{fig/cont_AB_Aur_1chan.pdf}
    \end{minipage}%
    \hfill
    \begin{minipage}[c]{0.75\linewidth}
        \centering
        \includegraphics[width=0.32\linewidth]{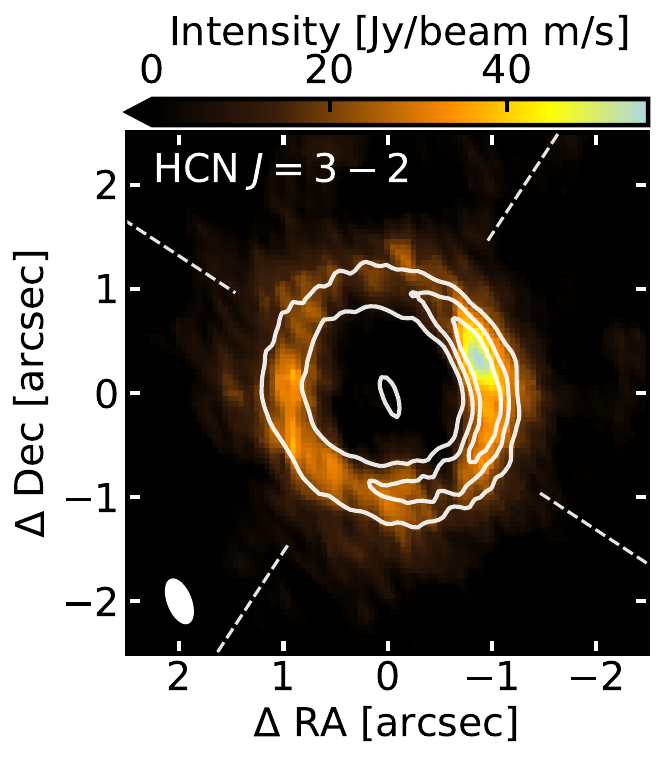}
        \includegraphics[width=0.32\linewidth]{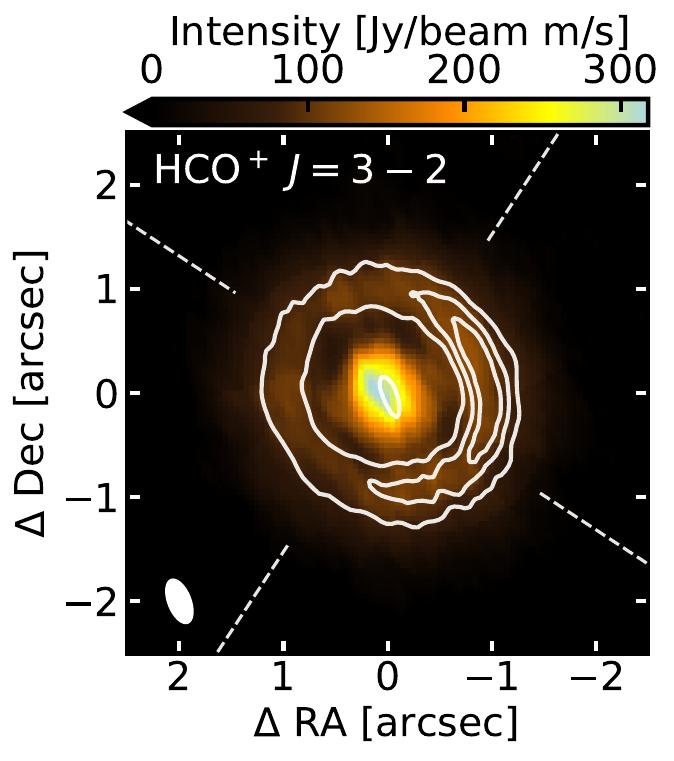}
        \includegraphics[width=0.32\linewidth]{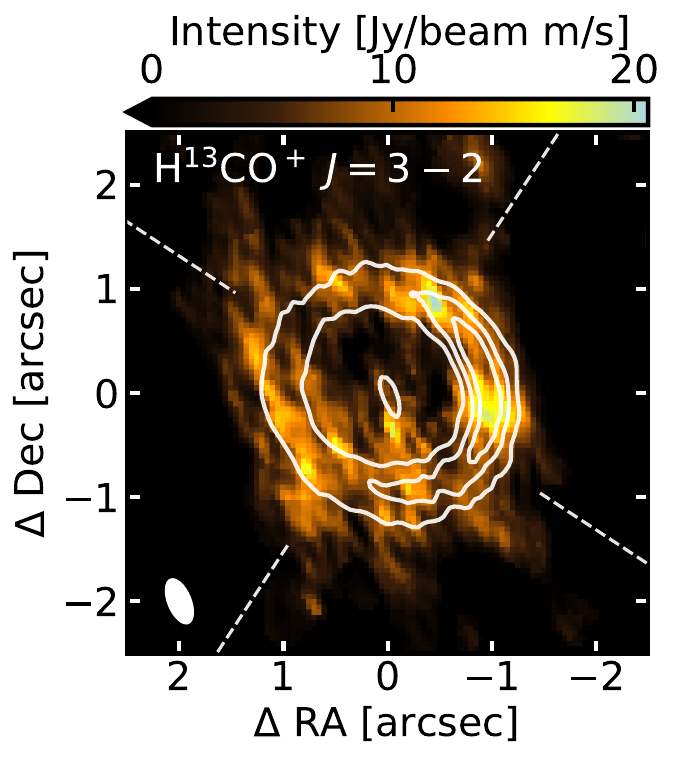}
        \includegraphics[width=0.32\linewidth]{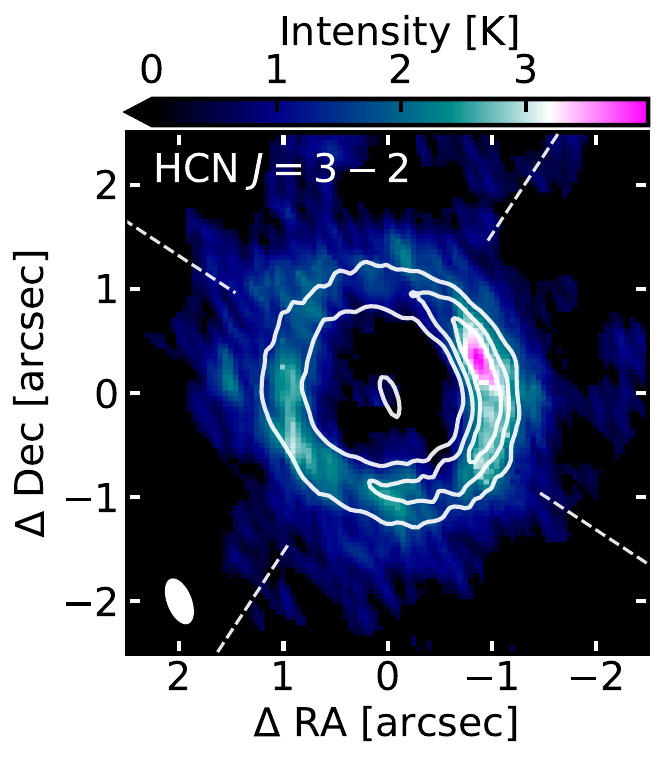}
        \includegraphics[width=0.32\linewidth]{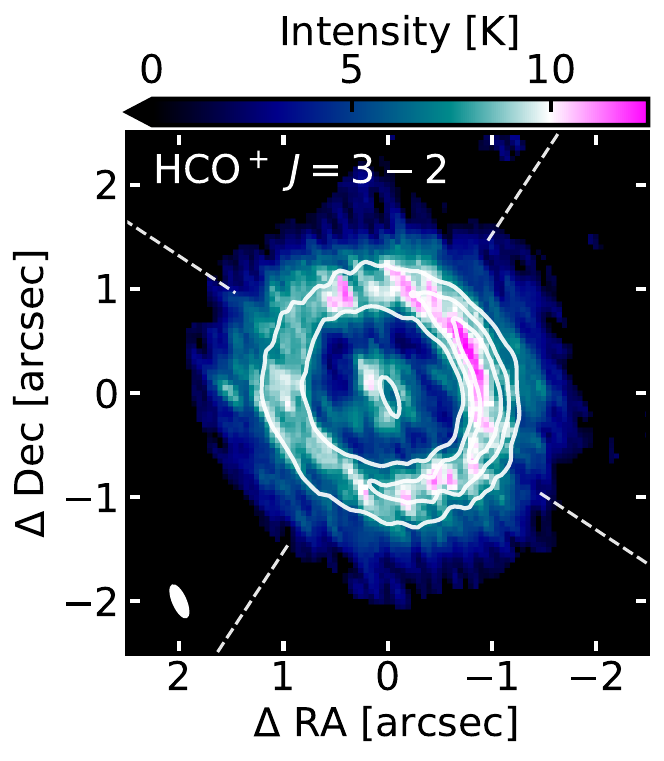}
        \includegraphics[width=0.32\linewidth]{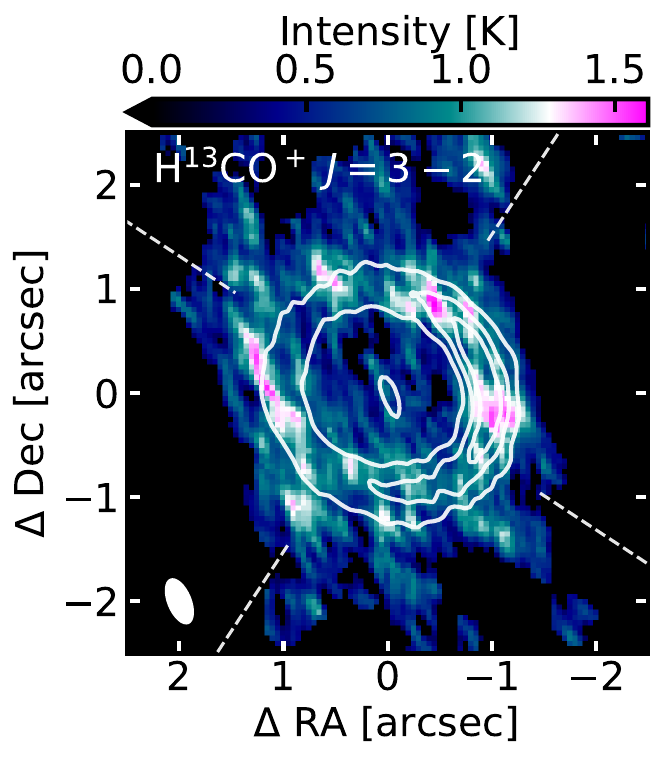}
    \end{minipage}
    \caption{Integrated intensity ({\bf Top}) and peak intensity ({\bf Bottom}) maps of HCN $J = 3-2$, HCO$^+$ $J = 3-2$ and H$^{13}$CO$^+$ $J=3-2$. Contours and annotations are identical to \fg{SO_C2H}. The HCO$^+$ maps shown are imaged with r $=0.5$ for better angular resolution. Unlike CS, SO and C$_2$H, whose ring peaks are beyond the continuum rim, HCN and HCO$^+$ align with the continuum ring and peak at the continuum dust overdensity.}
    \label{fig:HCN_HCOp}
\end{figure*}

Taken together, the distinct behavior of SO, C$_2$H, CS, HCN, and HCO$^+$ indicates that the azimuthal structure is chemically selective rather than a simple reflection of gas surface density variations.

\subsection{Column density estimates under LTE}\label{sec:columndensity}

To assess whether the observed intensity contrasts reflect variations in abundance, we derive apparent column densities under the assumptions of local thermodynamic equilibrium (LTE) and optically thin emission. Under the optically thin assumption, the velocity-integrated intensity $\int I_\nu dv$ is

\begin{equation}
\int I_\nu \, dv =
\frac{h c}{4\pi}A_{ul} N_u
,
\end{equation}
where $A_{ul}$ is the Einstein coefficient and $N_u$ is the column density in the upper state. The upper state density can be further related to the total column density $N$ by assuming LTE. Thus following the Boltzmann equation, we have \citep[e.g.,][]{Tielens2021}

\begin{equation}
\frac{N_u}{N} =
\frac{g_u}{Q(T_{\rm rot})}\exp\!\left(-\frac{E_u}{k_B T_{\rm rot}}\right)
,
\end{equation}
where $g_u$ is the upper-state degeneracy, $E_u$ is the upper-state energy, and $Q(T_{\rm rot})$ is the partition function at rotational temperature $T_{\rm rot}$. The total column density $N$ can be written as
\begin{equation}
N =
\frac{4\pi}{A_{ul} h c}
\frac{Q(T_{\rm rot})}{g_u}
\exp\!\left(\frac{E_u}{k_B T_{\rm rot}}\right)
\int I_\nu \, dv.
\end{equation}

For our analysis, the integrated intensity is derived from primary-beam-corrected moment~0 maps and converted to surface brightness using the synthesized beam solid angle. Spectroscopic parameters are adopted from CDMS \citep{MuellerEtal2001} and can be found in \tb{transition} and \tb{transition_archive}.

The peak brightness temperatures of CS, SO, and C$_2$H are overall $\lesssim3$\,K, well below the expected gas kinetic temperatures in the molecular layer, or even the midplane temperature of a Herbig disk \citep[e.g., HD~163296, $>20$\,K][]{DullemondEtal2020}. However, low brightness temperatures alone do not establish that the emission is optically thin, because spatial and spectral dilution may be significant. The thermal-linewidth opacity analysis presented below confirms that the SO transitions are optically thin, with $\tau_0\lesssim0.05$ for the disk-averaged fitting results. In contrast, the CS transitions have moderate disk-averaged opacities and may reach $\tau_0\sim0.8$--1.6 near the peak of the CS ring. We therefore treat the CS column densities derived under the optically thin assumption as apparent values or lower limits.

\subsubsection{SO and CS rotational analysis}\label{sec:rotation}

For SO, three detected transitions ($J_N=6_5-5_4$ from \citet{DutreyEtal2024,SpeedieEtal2025}, $5_6-4_5$, and $7_6-6_5$) spanning $E_u \sim 31$--51\,K allow a simultaneous determination of rotational temperature and column density as a function of radius. At each radial bin, we fit the three integrated intensities with a forward LTE model that predicts beam-averaged surface brightness for a given parameter pair $(N_{\rm SO}, T_{\rm rot})$.

The fitting is performed independently at each radius using a bounded least-squares minimization, with physically motivated limits of
$10^{12}$–$10^{15}$\,cm$^{-2}$ in $N_{\rm SO}$ and 5–200\,K in $T_{\rm rot}$. Each radius bin spans $0.1\arcsec$, corresponding to 15~au. 
To ensure radial continuity and numerical stability, the best-fit solution at one radius is used as the initial guess for the next radial bin. 
Uncertainties are estimated from the covariance matrix of the fit and propagated to derived quantities.

The resulting radial profiles (\fg{SO_N_T_r}) show that the SO-bright northern sector exhibits both higher apparent SO column densities and systematically elevated rotational temperatures (by approximately a factor of two) relative to the opposite side of the disk. In other words, SO is both more abundant and warmer on the blue-shifted NE side. 

Interestingly, this finding is the opposite of the temperature at the midplane of the continuum ring. Using C$^{17}$O hyperfine structure, \citet{DutreyEtal2024} determined lower temperature limits of 20~K at the continuum ring, and 40~K at the dust asymmetry, i.e., the dust is likely hotter on the west side than the east side. One plausible explanation is that C$^{17}$O is tracing the midplane temperature at the ring, while SO emission extends radially beyond the continuum and, likely, vertically above the midplane. We discuss a potential infall-related origin of the temperature enhancement in \Se{infall}, which might produce SO in the hotter upper layers of the disk. 

\begin{figure}
    \centering
    \includegraphics[width=.99\linewidth]{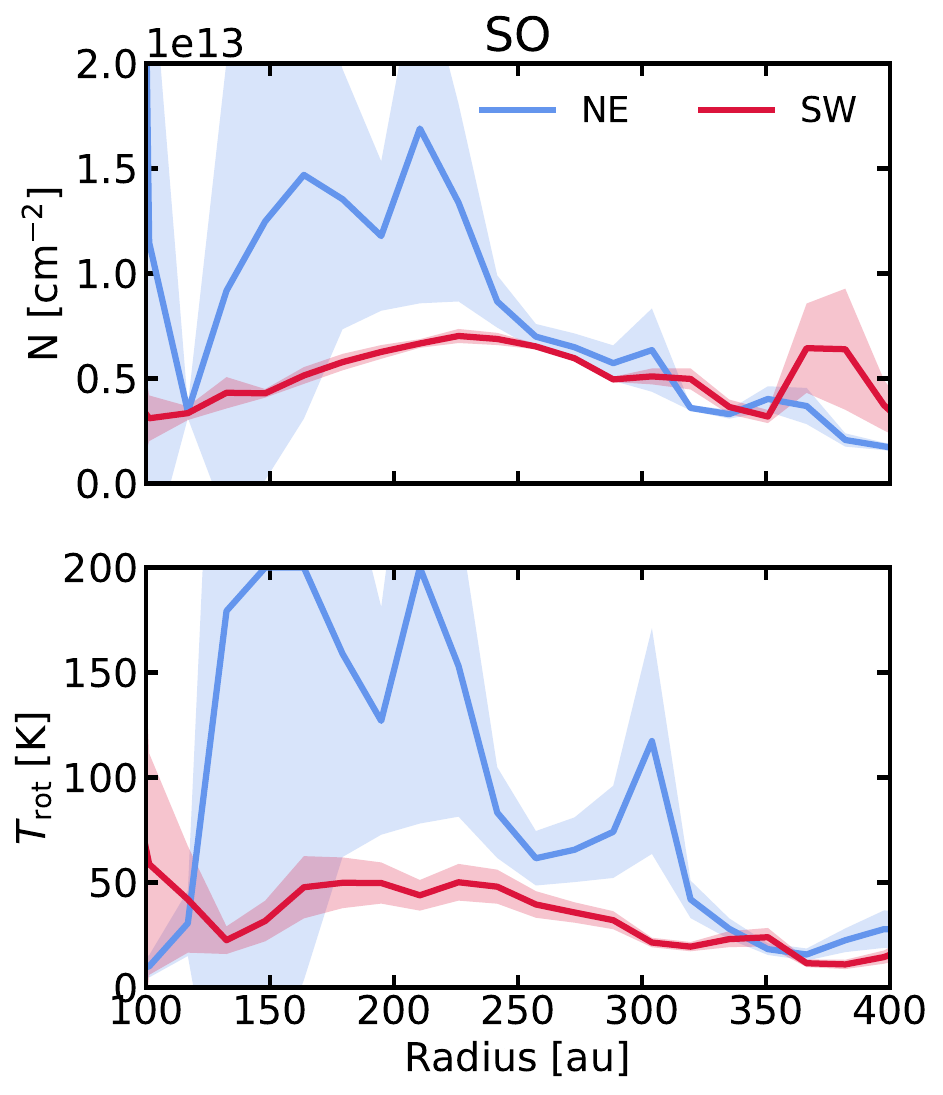}
    \caption{Rotational diagram fitting of SO based on three transitions (SO $J_N = 5_6$–$4_5$, and $7_6$–$6_5$ from this work, and SO $J_N = 6_5$–$5_4$ from \citet{SpeedieEtal2025}) spanning $E_u \sim 31$–51 K. The rotational temperature and column density are derived at each radius through a joint, robust least-squares fit to the three beam-convolved line intensities. {\bf Upper:} radial profile of the inferred SO column density. {\bf Lower:} radial profile of the rotational temperature. The shaded regions denote the 16th–84th percentile range of the posterior-equivalent uncertainties derived from the local covariance estimate.}
    \label{fig:SO_N_T_r}
\end{figure}

Similarly, three CS transitions ($J_N = 5$–$4$ from this work, $J_N = 3$–$2$ from \citet{Rivi`ere-MarichalarEtal2026}, and $J_N = 7$–$6$ from \citet{BoothEtal2026}) are available. We perform a CS rotational diagram analysis similar to that of SO. And the results are shown in \fg{CS_N_T_r}.

The rotational diagram confirms that CS is almost perfectly symmetric, in terms of both column densities and rotational temperature. The rotational temperature reaches a maximum of $\sim20$~K at the CS ring peak of $\sim210$~au. The values are sufficiently different from those derived based on SO transitions, even if they share similar radial extent, suggesting SO emission might originate from a vertically higher and warmer layer than CS and/or trace a distinct warm chemical component than the disk.

\begin{figure}
    \centering
    \includegraphics[width=.99\linewidth]{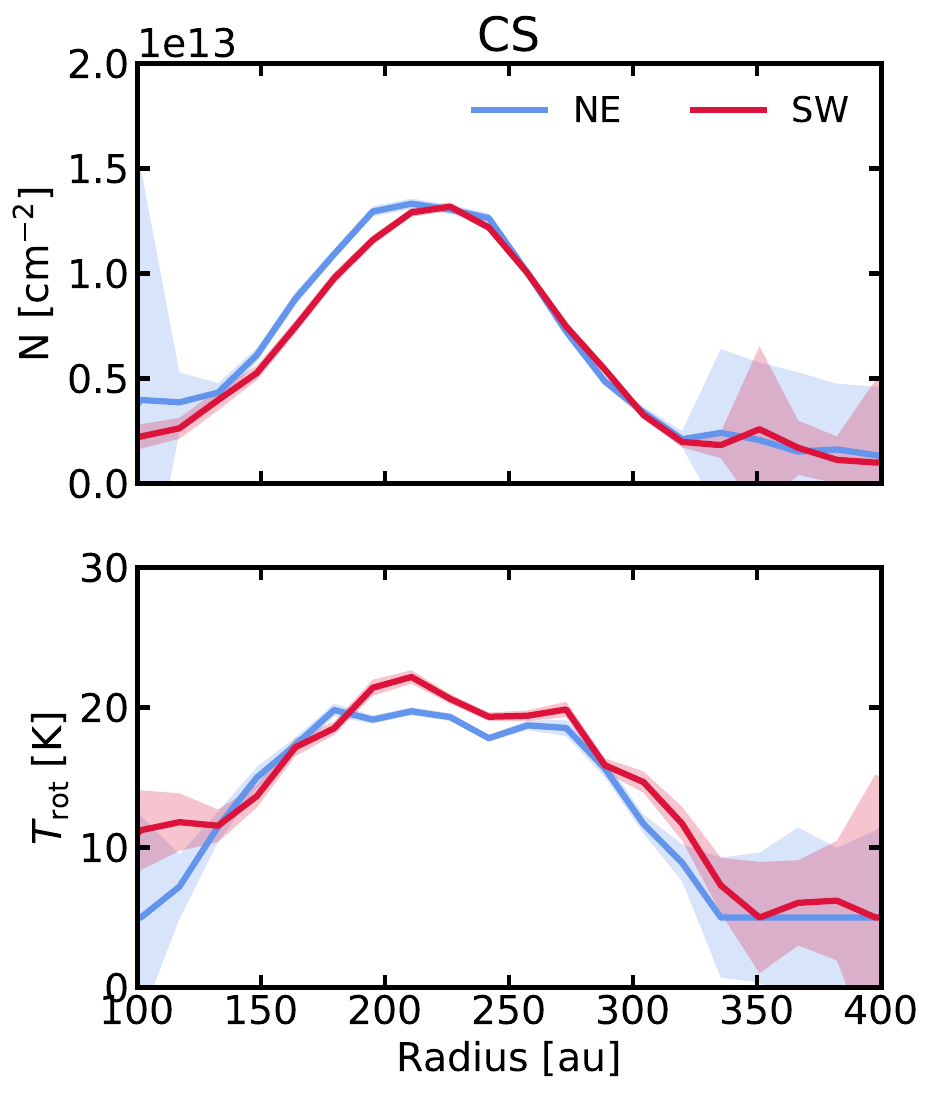}
    \caption{Rotational diagram fitting of CS based on three transitions (CS $J_N = 5$–$4$ from this work, CS $J_N = 3$–$2$ from \citet{Rivi`ere-MarichalarEtal2026}, and CS $J_N = 7$–$6$ from \citet{BoothEtal2026}) spanning $E_u \sim 14$–66 K. The rotational temperature and column density are derived at each radius through a joint, robust least-squares fit to the three beam-convolved line intensities. {\bf Upper:} radial profile of the inferred CS column density. {\bf Lower:} radial profile of the rotational temperature. The shaded regions denote the 16th–84th percentile range of the posterior-equivalent uncertainties derived from the local covariance estimate.}
    \label{fig:CS_N_T_r}
\end{figure}

For comparison, we also present the disk-averaged rotational diagram fits for CS and SO in \fg{rot_tot}. The total integrated line fluxes used in the fits are listed in \tb{transition} and \tb{transition_archive}. For both CS and SO, we assume that the emission originates from an elliptical region enclosed within a radius of 2.5\arcsec. We note that the assumed emitting area affects only the inferred disk-averaged column density and has no impact on the derived rotational temperature.

Qualitatively, the SO emission exhibits a rotational temperature approximately twice that of CS, consistent with the spatially resolved analysis. The disk-averaged rotational temperature of CS agrees remarkably well with the spatially resolved results presented in \fg{CS_N_T_r}. Although the disk-averaged rotational temperature of SO is lower than the peak value inferred from the spatially resolved analysis, the latter is subject to relatively large uncertainties.

As a consistency check on the optically thin assumption, we estimate the line-center optical depths using purely thermal intrinsic linewidths. Assuming that the kinetic temperature is equal to the fitted rotational temperature, the thermal FWHM is
\begin{equation}
\Delta v_{\rm th}
=
\sqrt{
\frac{8\ln 2\,k_{\rm B}T_{\rm rot}}{m_{\rm mol}}
},
\end{equation}
where $m_{\rm mol}$ is the molecular mass. For a Gaussian line profile, the line-center optical depth under LTE is \citep[cf. Eq. (11) of][]{LoomisEtal2018b}
\begin{equation}
\tau_0 =
\frac{c^3 A_{ul}}{8\pi\nu^3}
\frac{N_{\rm tot}g_u}{Q(T_{\rm rot})}
\exp\left(-\frac{E_u}{k_{\rm B}T_{\rm rot}}\right)
\left[
\exp\left(\frac{h\nu}{k_{\rm B}T_{\rm rot}}\right)-1
\right]
\left(\frac{4\ln 2}{\pi}\right)^{1/2}
\frac{1}{\Delta v_{\rm th}}.
\end{equation}
Using the disk-averaged SO fitting results, $T_{\rm rot}=37.4\pm7.2$\,K and $N_{\rm tot}=(4.99\pm1.34)\times10^{12}$\,cm$^{-2}$, we obtain a thermal FWHM of $0.190$\,km\,s$^{-1}$. The corresponding line-center optical depths are $\tau_0=0.051$, $0.032$, and $0.052$ for SO $J_N=6_5$--$5_4$, $5_6$--$4_5$, and $7_6$--$6_5$, respectively. All three SO transitions are therefore safely optically thin under the thermal-linewidth assumption.

For CS, the disk-averaged values $T_{\rm rot}=20.8\pm0.8$\,K and $N_{\rm tot}=(4.91\pm0.31)\times10^{12}$\,cm$^{-2}$ give a thermal FWHM of $0.148$\,km\,s$^{-1}$. We obtain $\tau_0=0.51$, $0.57$, and $0.30$ for CS $J=3$--$2$, $5$--$4$, and $7$--$6$, respectively. Because the CS emission is concentrated in a ring, the disk-averaged column density dilutes the higher local column densities and may underestimate the opacity near the emission peak. We therefore repeat the calculation using the peak values from the spatially resolved fit, resulting optical depths are $\tau_0=1.46$, $1.59$, and $0.77$ for CS $J=3$--$2$, $5$--$4$, and $7$--$6$, respectively. The first two transitions may therefore be optically thick near the peak of the CS ring, while $J=7$--$6$ may have moderate opacity. Consequently, although the disk-averaged emission has $\tau_0<1$, the optically thin approximation is not necessarily valid locally. The corresponding opacity correction factors are approximately 1.90, 2.00, and 1.44 for CS $J=3$--$2$, $5$--$4$, and $7$--$6$, respectively. This suggests that the optically thin approximation may underestimate the local CS column density by approximately 40--100\% near the peak of the ring.

The adopted thermal linewidths are smaller than our $0.4$\,km\,s$^{-1}$ channel spacing and therefore cannot be measured directly from these observations. Non-thermal broadening and Keplerian velocity gradients would reduce the effective line-center opacity for a fixed column density, whereas unresolved spatial beam dilution could imply a larger source-averaged column. The values above should therefore be regarded as consistency checks based on the adopted beam-averaged LTE model, rather than direct measurements of the intrinsic optical depths. We consequently retain the optically thin interpretation for SO, while treating the CS column densities as apparent values that may be subject to moderate opacity corrections.

The disk-averaged CS rotational temperature ($T_{\rm rot}$) and total column density ($N_{\rm tot}$) are comparable to those measured in the other two Herbig disks for which disk-averaged CS rotational analyses are available (HD~163296 and MWC~480; \citealt{LawEtal2025,LawEtal2026}). Relative to the sample of Sun-like disks analyzed by \citet{LawEtal2026}, the Herbig disks exhibit systematically lower CS column densities but higher rotational temperatures. This trend further supports the interpretation proposed by \citet{LawEtal2026}: the stronger UV radiation fields around Herbig stars may reduce the total CS column density through enhanced photodissociation, and/or the higher disk temperatures may suppress CO freeze-out, preserving more CO in the gas phase \citep{MiotelloEtal2023,TrapmanEtal2025}. Reduced CO depletion would in turn leave less elemental carbon available for the formation of CS.

Our fitting results differ from those reported by \citet{DutreyEtal2024}, who derived a rotational temperature of $\sim20$~K for SO using a parameterized ring model. The primary source of this discrepancy is likely from the data themselves: for the same SO transitions, our integrated fluxes are generally a factor of 2–3 higher than theirs. This difference is substantially larger than the typical absolute flux calibration uncertainty of $\sim10\%$ for both NOEMA and ALMA, suggesting that calibration uncertainties alone cannot account for the discrepancy.

One additional clue is that \citet{DutreyEtal2024} measured the SO $J_N=5_6-4_5$ and $J_N=7_6-6_5$ line fluxes from a stacked data cube, which may have affected the recovered fluxes. However, we note that the SO $J_N=5_6-4_5$ observations analyzed in both studies originate from the same ALMA program (2021.1.00690.S; PI: R.~Dong), although the data cube used in this work was independently reduced by \citet{SpeedieEtal2025}.

Investigating the origin of these discrepancies is beyond the scope of this work, and we therefore refrain from drawing conclusions regarding the differences with \citet{DutreyEtal2024}. We emphasize that the CS and SO rotational diagram analyses presented here were performed using a consistent methodology. We note, however, that the some transitions included in our analysis were compiled from archival observations obtained through different observing programs, see \App{archival_data}, which may introduce additional systematic uncertainties.

\begin{figure}
    \centering
    \includegraphics[width=1.\linewidth]{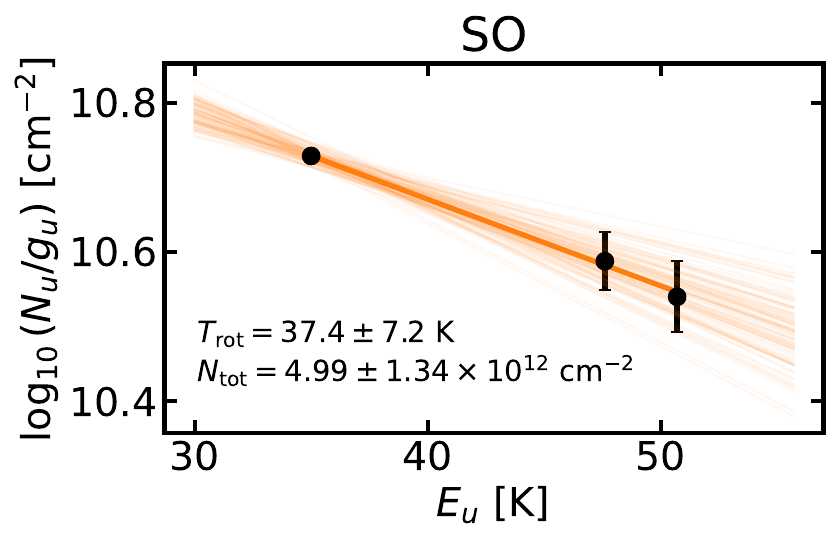}
    \includegraphics[width=1.\linewidth]{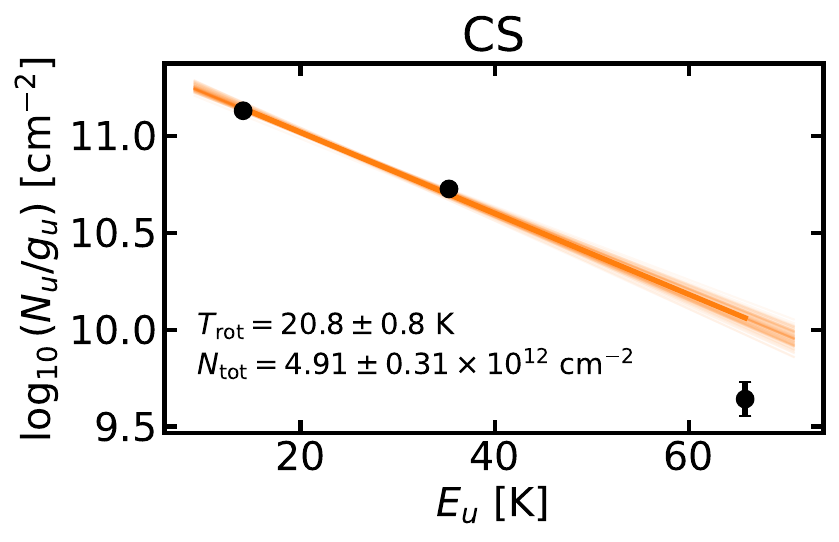}
    \caption{Rotational diagram fitting for CS and SO, but using disk-averaged brightness. Orange lines show random draws from the fit posteriors.}
    \label{fig:rot_tot}
\end{figure}

\subsubsection{C$_2$H column densities}

For C$_2$H, only one transition is available, preventing an independent temperature determination. We therefore compute column densities over a range of assumed rotational temperatures between 15 and 160\,K. 

The inferred column densities depend on the assumed $T_{\rm rot}$. In \fg{N_peak_T_rot}, we show the derived column density of C$_2$H at the ring peak of their emission. The C$_2$H column density reaches a minimum when the assumed $T_{\rm rot}$ is equal to $E_u = 25$~K for the C$_2$H $N=3-2$ hyperfine structure group. 

Although the derived column densities vary with $T_{\rm rot}$, their values are always around $1-3\times10^{13}$~cm$^{-2}$ within the reasonable range of $T_{\rm rot}$. We also mark the peak column densities of CS and SO derived from the rotational diagram at both sides in \fg{N_peak_T_rot}. 

\begin{figure}
    \centering
    \includegraphics[width=\linewidth]{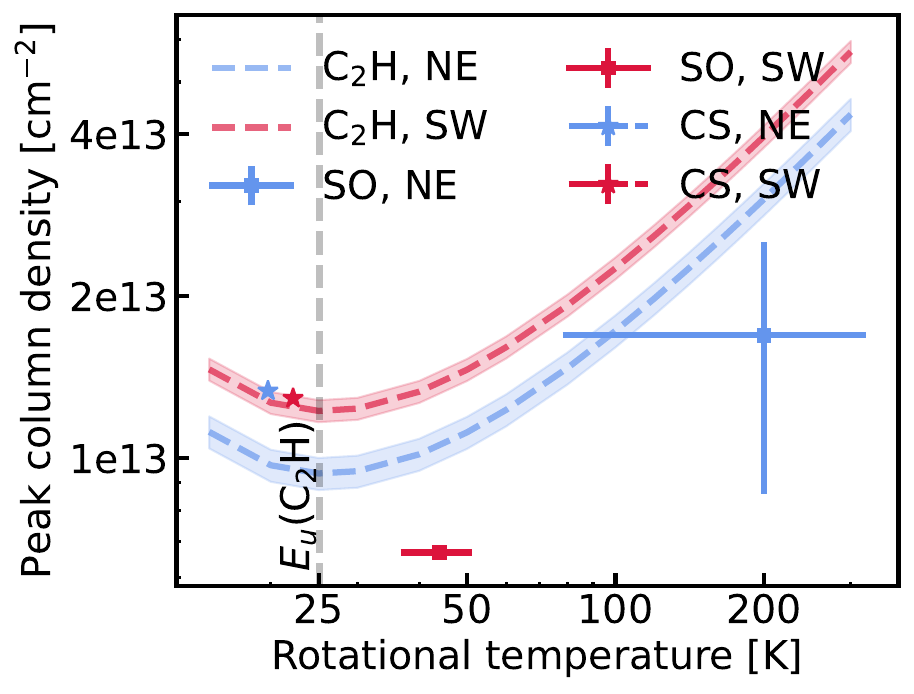}
    \caption{Peak column densities of C$_2$H (dashed lines) as a function of assumed rotational temperature. The dashed vertical line marks the upper-state energy $E_u$ for the C$_2$H $N=3-2$ transitions. The red color represents the values averaged along the SW side, and the blue color represents the values averaged along the NE side of the disk. The peak column densities and rotational temperatures of CS (stars, \fg{CS_N_T_r}) and SO (squares, \fg{SO_N_T_r}) from rotational diagram analysis are shown for comparison. }
    \label{fig:N_peak_T_rot}
\end{figure}

Across the explored temperature range, C$_2$H shows a persistent south-to-north contrast unless substantial temperature differences are assumed. We discuss the possibility that the difference is purely due to the excitation temperature difference in the next section.

\subsection{Excitation versus abundance effects}\label{sec:excitation}

As noted in \fg{N_peak_T_rot}, due to the low upper state energy of the C$_2$H $N=3-2$ transitions, a higher rotational temperature will imply a higher intrinsic column density for emission with the same intensity. In other words, even if the column densities are the same on both sides of the disk, the line emission intensity could still be fainter at the warmer NE side because of a difference in rotational temperature. 

The SO rotational analysis demonstrates that the northern sector is indeed potentially warmer. If other species share a similar temperature contrast, part of their intensity variation could be excitation-driven.

To quantify this effect for C$_2$H, we adopt a representative temperature contrast motivated by the SO results (e.g., $T_{\rm rot}=100$\,K in the north and 60\,K in the south) and recompute the column densities. Under this assumption, the inferred C$_2$H column densities become more symmetric (\fg{C2H_100_60}), indicating that excitation differences can account for a significant fraction of the observed C$_2$H intensity contrast.

\begin{figure*}
    \centering
    \includegraphics[width=.9\linewidth]{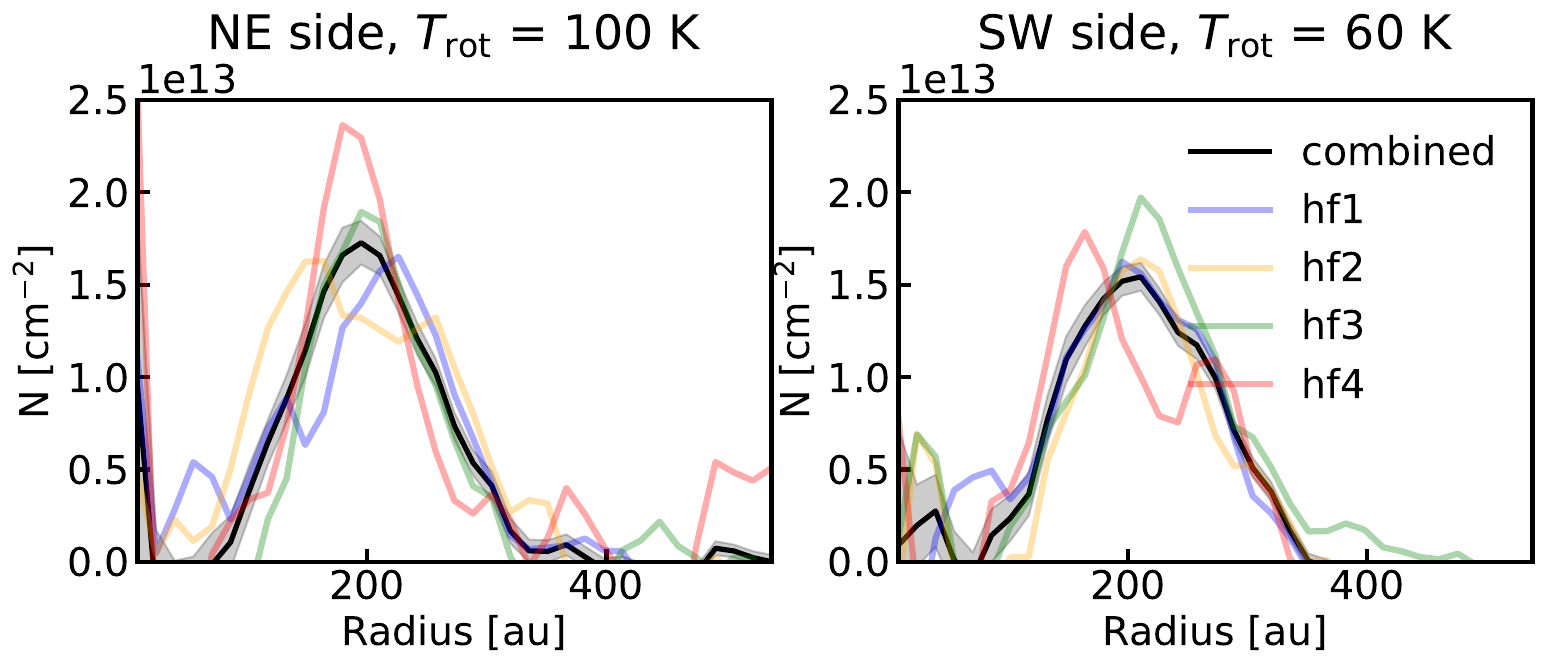}
    \caption{Derived C$_2$H column densities on the northern and southern sides of the disk assuming a temperature contrast motivated by the SO rotational analysis, i.e., $T_{\rm rot}=100$\,K in the north and 60\,K in the south.}
    \label{fig:C2H_100_60}
\end{figure*}

Yet, excitation differences alone cannot reproduce the full set of observational constraints. Applying the same temperature contrast to CS and SO implies that CS should also become asymmetric at a detectable level. This is inconsistent with the observed axisymmetric CS ring and the results of the CS rotational diagram analysis. In addition, the SO enhancement remains pronounced even after accounting for the higher $T_{\rm rot}$ in the north, requiring a genuine increase in SO column density. These tensions are illustrated in \fg{CS_SO_100_60}.

\begin{figure*}
    \centering
    \includegraphics[width=.9\linewidth]{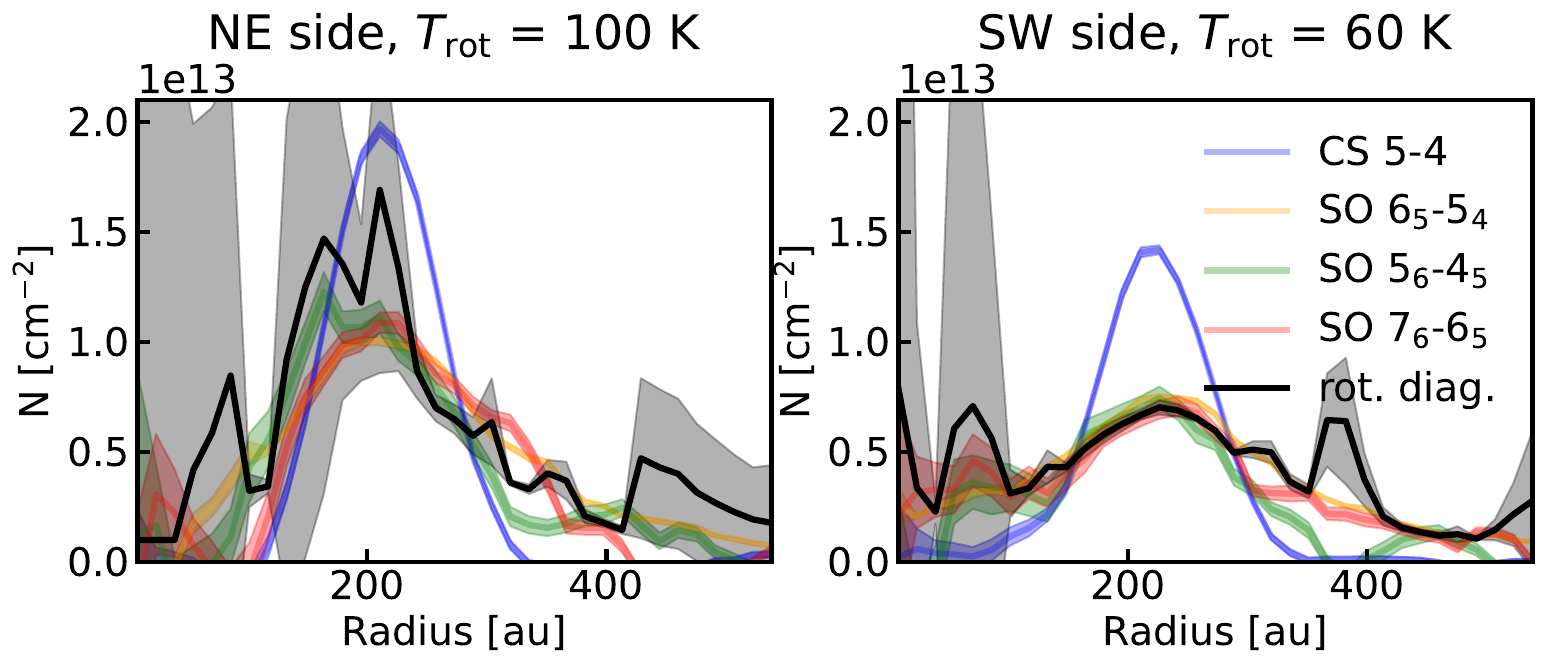}
    \caption{Derived CS and SO column densities on the northern and southern sides of the disk under the same temperature contrast assumption as in \fg{C2H_100_60}. While the C$_2$H column density contrast can be reduced by temperature effects, the SO enhancement and the near-axisymmetric CS distribution cannot be simultaneously reproduced by excitation differences alone.}
    \label{fig:CS_SO_100_60}
\end{figure*}

As we noted for the comparison of rotational temperatures between CS and SO, the emission layers of C$_2$H, SO, and CS are not necessarily co-spatial, so their rotational temperatures need not be identical. Quantifying the full 2D temperature and abundance structure would require dedicated multi-transition observations and radiative-transfer modeling \citep[e.g.,][]{LawEtal2021b}. Nevertheless, the combination of (i) a strong SO enhancement in the north, (ii) a C$_2$H enhancement in the south, and (iii) an approximately symmetric CS ring indicates that excitation differences contribute to, but do not fully explain, the observed azimuthal chemical differentiation.

\section{Physical-chemical model of the AB~Aur disk}\label{sec:model}

SO, CS, and C$_2$H respond differently to variations in elemental C/O and to changes in temperature and ionization conditions \citep[e.g.,][]{LeGalEtal2021,KeyteEtal2023,BosmanEtal2021b,ZagariaEtal2025}. In general terms:

\begin{itemize}
\item SO is favored in relatively O-rich environments and can be enhanced by thermal desorption or shock-driven release of O-bearing species.
\item CS tends to be less sensitive to modest C/O variations and often traces the overall gas distribution.
\item C$_2$H is commonly associated with elevated C/O or hydrocarbon-rich chemistry in UV-irradiated disk layers.
\end{itemize}

The observed pattern in AB~Aur, SO-bright in the north, C$_2$H-bright in the south, and approximately symmetric CS, therefore qualitatively suggests that the northern sector is comparatively O-rich, while the southern sector hosts more hydrocarbon-favored chemistry.

To place this interpretation on firmer footing, we compare the inferred column densities and CS/SO ratios to a grid of gas-grain chemical models with varying elemental C/O ratios, using the \texttt{ALCHEMIC} chemical modeling code \citep{SemenovEtal2010,Semenov2017,SemenovEtal2018}. The goal of this modeling is not to reproduce the full spatial structure of AB~Aur, but to explore how variations in elemental C/O and local physical conditions affect the relative abundances of SO, CS, and C$_2$H at radii comparable to the observed ring ($\sim150$–220\,au).

\subsection{Disk physical structure}

We use a 1+1D parametric disk physical structure based on \citet{Riviere-MarichalarEtal2020,Riviere-MarichalarEtal2022,Rivi`ere-MarichalarEtal2026} that qualitatively fits the various observed molecular emission data (e.g., CS, CO, etc.) at a lower angular resolution of $\sim 1\arcsec$. 

The disk structure is assumed to be azimuthally symmetric. The gas temperature in the vertical direction is modeled by interpolating between the midplane $T_{\rm mid}$ and atmospheric $T_{\rm atm}$ temperatures, whereas both $T_{\rm mid}$ and $T_{\rm atm}$ are described by the radial power law:
\begin{align}
T_{\rm mid}(r) &= T_{\rm mid,0}\left(\frac{r}{R_0}\right)^{-0.1},\\
T_{\rm atm}(r) &= T_{\rm atm,0}\left(\frac{r}{R_0}\right)^{-0.1}.
\end{align}
Here, $T_{\rm mid,0}=42~{\rm K}$, $T_{\rm atm,0}=70~{\rm K}$, and the characteristic radius $R_0$ is 98~au.
The vertical temperature profile for the vertical heights $z$ below 4 pressure scale heights $H(r)$ is described by the following Equation:
\begin{equation}
T_{\rm g}(r,z) = T_{\rm mid}(r) + \left[T_{\rm atm}(r)-T_{\rm mid}(r)\right] \sin^4\!\left(\frac{\pi z}{2 H(r)}\right),
\end{equation}
where the gas pressure scale height $H(r)$ is computed assuming hydrostatic equilibrium:
\begin{equation}
H(r) = \sqrt{\frac{k_{\rm B}\,T_{\rm mid}(r)\,(r\,{\rm})^{3}}
{\mu\,m_{\rm H}\,G\,M_\star}}.
\end{equation}
Here, following \citet{Riviere-MarichalarEtal2020}, we assume that the AB Aur stellar mass is $2.4M_\odot$, and the gas mean molecular weight $\mu = 2.4$. The $k_{\rm B}$ is the Boltzmann constant and $G$ is the gravitational constant. For for the disk heights $z\ge 4H(r)$, $T_{\rm g}(r,z)=T_{\rm atm}(r)$. The gas and dust temperatures are assumed to be equal.

The gas surface density profile is described by
\begin{equation}
\Sigma(r) = \Sigma_0\left(\frac{r}{R_0}\right)^{-3/2},
\end{equation}
where $\Sigma_0 = 0.5~{\rm g\,cm^{-2}}$ at $R_0=98$~au. The gas volume density is calculated from the gas surface density assuming hydrostatic equilibrium:
\begin{equation}
n_{\rm H}(r,z) = n_0(r)\exp\!\left(-\frac{z^2}{2H(r)^2}\right),
\end{equation}
where the midplane gas density is $n_0(r)$ derived from the total $\Sigma(r)$. The dust distribution is assumed to be co-spatial with the gas, with a gas-to-dust mass ratio of 40. The appropriate gas-to-dust mass ratio of AB~Aur remains uncertain from observational perspectives. Based on CO isotopologue observations, \citet{StapperEtal2024} estimated a total gas disk mass of $0.17\,M_\odot$. Combined with the dust mass inferred from the millimeter continuum, this would imply an extreme global gas-to-dust mass ratio of $\sim\!5000$, suggesting that the dust mass might be underestimated owing to optically thick continuum emission. However, the mm--cm SED modeling of \citet{Rivi`ere-MarichalarEtal2024} inferred a dust mass of only $\sim30\,M_\oplus$, merely a factor of $\sim3$ larger than that derived from the single-wavelength millimeter continuum, suggesting that optical depth alone is unlikely to fully reconcile the discrepancy. Conversely, chemical modeling by \citet{Riviere-MarichalarEtal2020} found that a reduced gas-to-dust mass ratio of 40 better reproduces the observed molecular abundances. Given these uncertainties and to facilitate comparison with previous studies, we adopt a fiducial gas-to-dust mass ratio of 40 throughout this work.

We used this parametrization to define the disk structure on a 1+1D $100\times100$ grid spanning radial distances from $r=1$ to $r=400$ au and vertical heights up to 120~au. The unattenuated FUV radiation field at each radius is computed by adding the stellar UV flux $\chi_*(r)=12\,000 \chi_{0} \times (r_{\rm au}/100)^{-2}$
and interstellar UV flux $\chi_{\rm ISM}=1\chi_{0}$, where $\chi_{0}$ is in the \citet{Draine1978} units. After that, for each grid cell, the local FUV RF is calculated by taking into account extinction in the vertical direction and the direction toward the star by the ISM-like dust grains. 

Using the total stellar X-ray luminosity of $4\times10^{29}$~erg\,s$^{-1}$ \citep{TelleschiEtal2007}, the local X-ray ionization rate is calculated following Eq.~(37) from \citet[][see their Section 3.3.2]{Armitage2015}, assuming the solar metallicities, with the average X-ray photon energy of 3~keV, $\zeta_{X,1} = 6\times 10^{-12}$~s$^{-1}$, $\zeta_{X,2} = 10^{-15}$~s$^{-1}$, $\Sigma_{X,1} = 3.4\times 10^{-3}$~g~cm$^{-2}$, $\Sigma_{X,2} = 1.59$~g~cm$^{-2}$, $a = 0.4$, and $b=0.65$. The unattenuated cosmic ray (CRP) ionization rate is assumed to be $\zeta_{\rm CR}=1.3\times10^{-17}$~s$^{-1}$. The local CRP ionization rate in each grid cell is calculated by rescaling the unattenuated rate by the scaling factor between the surface density upward from the grid cell and the critical surface density after which the CRP rate starts to decline, $\exp(\Sigma(r,z)/\Sigma_{\rm CRP})$, where $\Sigma_{\rm CRP} = 96$~g~cm$^{-2}$.
Ionization due to the decay of short-lived radionuclide (SLR) is also taken into account, with the SLR ionization rate $\zeta_{\rm SLR}=6.5\times10^{-19}$~s$^{-1}$.

The adopted chemical model is from \citet[][]{SemenovEtal2018} and is based on the public Kinetic Database for Astrochemistry (KIDA) network \citep{WakelamEtal2015,MolyarovaEtal2017}. The self-shielding of H$_2$ is calculated by Eq.~(37) from \citet{DraineBertoldi1996}. The mutual shielding of CO by H$_2$ and dust, and the CO self-shielding are calculated from a precomputed table of \citet[][Table~11]{LeeEtal1996}. Adsorption with 100\% efficiency, thermal and non-thermal desorption of neutral species and electrons, and surface reactions in the several uppermost ice monolayers are taken into account in the two-phase formalism.  The desorption processes include thermal, CRP-driven, UV-driven, and chemical desorption with an efficiency of 1\%. The UV photodesorption yield of $10^{-5}$ was assumed. The representative dust grains for chemical modeling are uniform amorphous silicate particles of olivine stoichiometry with a density of $3$~g\,cm$^{-3}$ and a radius of $0.1\,\mu$m. Each grain has $1.88\times10^6$ surface sites for surface recombination reactions, which are modeled by the Langmuir-Hinshelwood mechanism. 

The adopted chemical network comprises approximately 650 species across 12 elements and includes more than 7\,800 reactions. The initial abundances are based on the ``low metals'' elemental abundances \citep{SemenovEtal2018}, with hydrogen mainly in molecular form and the standard C/O ratio of $\approx 0.44$ (see \tb{tabinit_abunds}). Using the adopted parametric disk and chemical kinetics model, we simulated the chemical evolution in the AB Aur disk over $t = 1$~Myr for C/O ratios ranging from $\approx 0.44$ to 2 (by scaling the initial oxygen abundance).

\begin{table}
\centering
\caption{Initial abundances for the ``standard'' solar C/O case.}
\begin{tabular}{ll|ll}
\hline\hline
Species & Relative abundances & Species & Relative abundances \\
\hline
H$_2$ & $0.499$               & H     & $2.00 \times 10^{-3}$ \\
He    & $9.75 \times 10^{-2}$ & C     & $7.86 \times 10^{-5}$ \\
N     & $2.47 \times 10^{-5}$ & O     & $1.80 \times 10^{-4}$ \\
S     & $9.14 \times 10^{-8}$ & Si    & $9.74 \times 10^{-9}$ \\
Na    & $2.25 \times 10^{-9}$ & Mg    & $1.09 \times 10^{-8}$ \\
Fe    & $2.74 \times 10^{-9}$ & P     & $2.16 \times 10^{-10}$\\
Cl    & $1.00 \times 10^{-9}$ &       &                       \\
\hline
\end{tabular}
\label{tab:tabinit_abunds}
\end{table}

\subsection{Model grid and comparison methodology}

We compare the model-predicted column density and CS/SO ratio with the azimuthally averaged values inferred separately for the northern and southern sectors. The purpose of this comparison is to assess which combinations of C/O and physical conditions can reproduce:

\begin{itemize}
\item Enhanced SO column density in the northern sector,
\item Enhanced C$_2$H in the southern sector,
\item Approximately symmetric CS distribution.
\end{itemize}

\begin{figure}
    \centering
    \includegraphics[width=\linewidth]{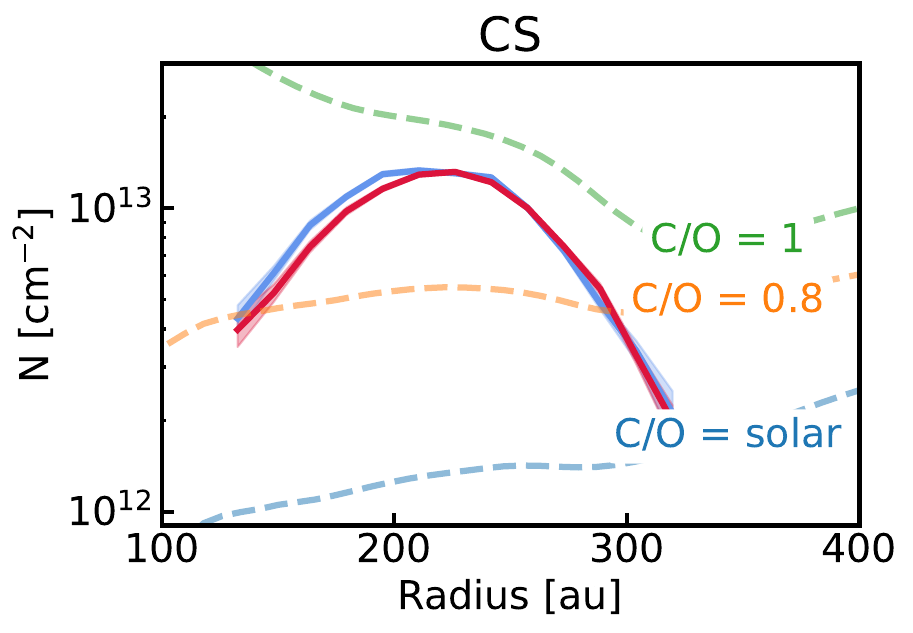}
    \includegraphics[width=\linewidth]{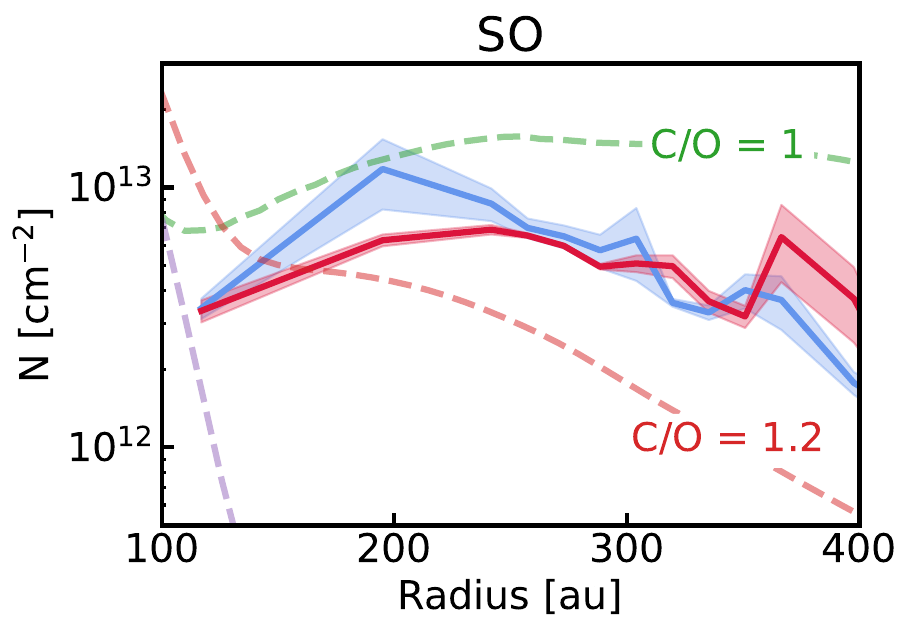}
    \includegraphics[width=\linewidth]{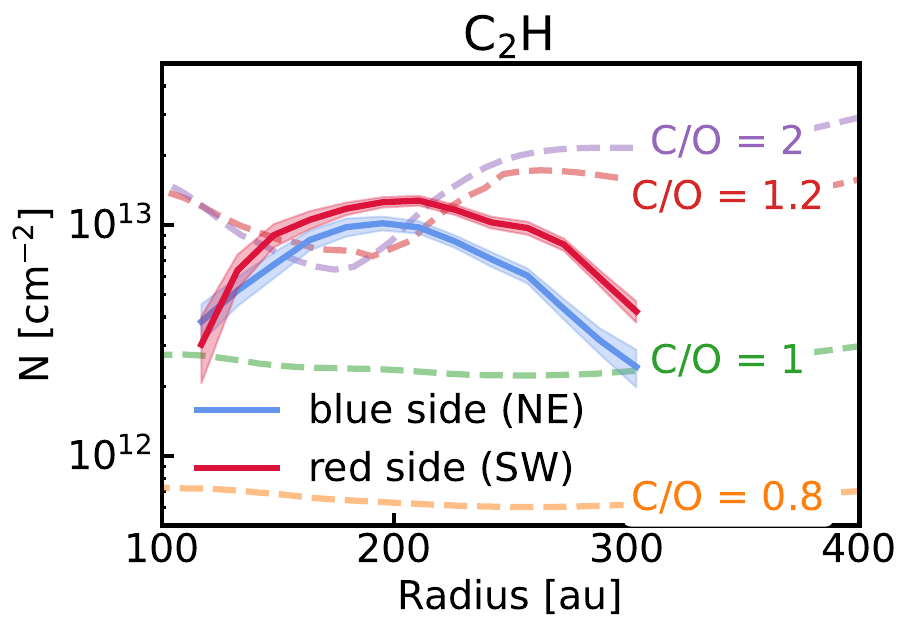}
    \caption{The observationally derived column density of CS, SO and C$_2$H compared to those of our chemical productions, when assuming $T_{\rm rot} = 40$~K for C$_2$H. For CS and SO, we used the column densities derived from the rotational diagram.}
    \label{fig:N_r}
\end{figure}

We compare the chemical predictions with the inferred column densities of CS, SO, and C$_2$H in \fg{N_r}. The rotational temperature used in this figure is fixed at $40$~K, corresponding to the midplane temperature in the disk model at $r=200$~au, where the molecular line emission peaks. 

The column densities show a clear C/O dependence, with CS and C$_2$H positively correlated with C/O, and SO negatively correlated with C/O. In addition, we also compare the observationally derived CS/SO ratio with the model outputs in \fg{N_CS_SO}. 

\begin{figure}
    \centering
    \includegraphics[width=\linewidth]{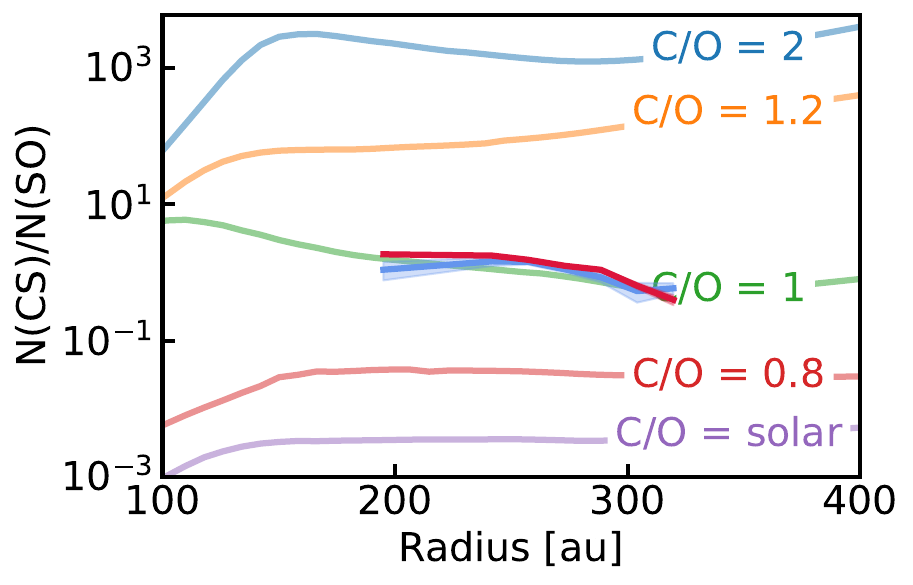}
    \caption{The observationally derived column density ratio between CS and SO compared to those of our chemical modeling results.}
    \label{fig:N_CS_SO}
\end{figure}

Although strong asymmetries are observed in the line emission morphology, the observationally derived abundances and predictions from contemporary chemical models cannot simultaneously reproduce the abundances of CS, SO, and C$_2$H in disks with a single C/O ratio. Specifically, CS prefers a model with the C/O ratio slightly below unity, SO prefers a model with the C/O ratio slightly above unity, and the CS/SO ratio points to a C/O ratio near unity. Furthermore, the high inferred abundance of C$_2$H alone indicates a substantially non-solar C/O ratio $\gg1$. This confirms earlier evidence that the current astrochemical networks are likely to be missing some sulfur chemistry pathways or species, such as sulfur allotropes or organosulfides, or detailed shock-induced processes \citep[e.g.][]{SemenovEtal2018,LeGalEtal2019a,Riviere-MarichalarEtal2022,FerrariEtal2024}. It is unlikely that the optically thin emission and local thermodynamic equilibrium assumptions used in deriving the column density are responsible for this disagreement, since several past studies of different disks with different chemical models and column density estimation methods reached the same conclusion. Therefore, by considering all the above constraints, we propose that the C/O ratio in the AB Aur disk is greater than or equal to 1; a solar/ISM-like C/O ratio is very unlikely \citep[see also][]{DutreyEtal2024,Rivi`ere-MarichalarEtal2026}. 

Moreover, the observed anti-correlation between SO and C$_2$H is qualitatively consistent with a scenario in which the northern sector has a lower effective C/O than the southern sector. Indeed, all diagnostics favor a higher C/O ratio on the red-shifted side of the disk than on the blue-shifted side, although the difference is small. Our objective is not to derive a unique chemical model of AB~Aur, but to investigate the differential response of the molecular abundances to changes in the gas-phase C/O ratio. While the predicted absolute abundances depend on uncertain physical parameters, including the gas-to-dust mass ratio, sulfur depletion, UV penetration, and the vertical temperature structure, the qualitative dependence of CS, SO, and C$_2$H on C/O is much less sensitive to these assumptions. The modeling, therefore, supports a relative difference in effective C/O between the two azimuthal sectors rather than providing a precise absolute measurement.

\section{Discussions}\label{sec:discussions}
\subsection{The origin of the molecular line azimuthal variation}\label{sec:origin}

The anti-correlation between SO and C$_2$H, together with the nearly axisymmetric CS distribution, indicates genuine azimuthal chemical differentiation in the AB~Aur disk. In this section, we explore two physically motivated scenarios that may account for the observed pattern: (i) infall-driven oxygen-rich chemistry in the northern sector, and (ii) carbon-enhanced chemistry associated with disk substructure and potential planet formation in the southern sector.

\subsubsection{Infall induced temperature elevation and O-rich chemistry}\label{sec:infall}

The spatial coincidence between the SO-bright region and the ``merging zone'' where the large-scale streamers intersect the outer disk \citep{DutreyEtal2024,SpeedieEtal2025} strongly suggests that late infall is actively reshaping the local chemistry in AB~Aur. In this picture, gas with approximately ISM-like composition (C/O$\sim 0.5$) is accreted along filamentary flows and impacts the disk surface at supersonic velocities. The resulting low- to intermediate-velocity shocks heat the gas and dust, desorb O-bearing ices (in particular H$_2$O and CO$_2$) from grain mantles, and drive a transient enhancement of gas-phase atomic oxygen and OH. Both the elevated gas temperature and the enhanced O reservoir efficiently channel sulfur into SO (and, to a lesser degree, SO$_2$) through reactions such as S + OH $\rightarrow$ SO + H and SH + O $\rightarrow$ SO + H, boosting the SO abundance in the post-shock gas \citep[e.g.,][]{LeGalEtal2021,KeyteEtal2023}. 

In this scenario, the northern side of the disk, where the streamer-disk interaction is strongest, is continuously replenished with O-rich material. This drives the local gas-phase C/O ratio back toward ISM-like values and maintains a high SO/CS ratio. The released energy can also locally elevate the disk temperature, at least in the surface layer. In contrast, the southern side of the disk remains relatively undisturbed by infall and retains the C-enriched composition built up through internal disk evolution (see protoplanet). Thus, the observed morphology is a natural outcome of spatially localized, time-dependent, non-azimuthally symmetric infall: the SO-bright, C$_2$H-faint region traces the O-rich, shock-processed gas at the infall impact site \citep{GarufiEtal2022a,HuangEtal2024,ZagariaEtal2025}, while the opposite side of the disk reflects the underlying C-rich disk chemistry.

The observed properties of the SO emission are qualitatively consistent with this picture. First, the bright, compact SO emission near the streamer impact region is indicative of elevated rotational temperatures and densities compared to the more extended CS ring. Second, the similar azimuthal morphology among multiple SO transitions spanning a range of upper-level energies ($J_N=6_5-5_4$, $5_6-4_5$, $7_6-6_5$) \citep{DutreyEtal2024,SpeedieEtal2025} suggests that the enhancement is primarily driven by abundance rather than solely by local excitation effects. Third, the nearly axisymmetric CS $J=5-4$ ring and the C$_2$H asymmetry peaking on the opposite side argue against a purely total-density-driven explanation: if the northern side simply hosted a higher gas surface density, CS would also be expected to exhibit a comparable asymmetry. Instead, the higher CS/SO ratio in the south suggests an azimuthal variation in the sulfur chemical network, linked to the underlying C/O ratio.

The timescales are also plausible. Hydrodynamical simulations of late infall onto evolved disks show that filamentary accretion can persist for $\gtrsim 10^5$~yr and can deliver a non-negligible fraction of the final disk mass \citep[e.g.,][]{KuffmeierEtal2020,KuffmeierEtal2023}. Chemical models of shock-processed gas indicate that SO enhancements can arise on timescales of $10^2$–$10^3$~yr and remain elevated for up to $10^4$–$10^5$~yr, depending on the post-shock density and ionization rate \citep[e.g.,][]{vanGelderEtal2021}. 

This interpretation leads to several testable predictions. If infall is responsible for the SO enhancement, other shock or streamer tracers, such as warm H$_2$CO \citep[e.g.,][]{KaufmanNeufeld1996,vanDishoeckEtal2021} and other ice-mantle products \citep[e.g., CH$_3$OH,][]{TychoniecEtal2021}, should also peak near the streamer-disk junction. A dedicated search of those molecules in AB~Aur would provide a smoking gun of infall-induced shock chemistry. In addition, we expect velocity gradients in SO that deviate subtly from purely Keplerian rotation, reflecting the inflow component of the gas motion. With our current spectral resolution ($\sim0.3$~km/s), we could not address this question. We, however, refer interested reader to \citet{SpeedieEtal2025} for a detailed analysis of SO $J_N=6_5-5_4$ in $\sim0.08$~km/s, that show a strong hint of this. Finally, other sulfur-bearing species that are more sensitive to high-temperature chemistry (e.g., SO$_2$, OCS) may be selectively enhanced in the infall-impacted region. Spatially resolved multi-line observations of these species, combined with detailed thermo-chemical modeling that includes a time-dependent infall boundary condition, are needed to quantify whether the observed SO brightness and inferred C/O contrast can be fully reproduced by late infall alone.

We note that the O-rich infall scenario does not preclude additional processes, such as local variations in dust properties or vertical mixing, from contributing to the observed asymmetry. However, any mechanism that aims to explain the SO–C$_2$H anti-correlation must simultaneously deliver a localized enhancement of O-bearing volatiles and avoid producing a comparable enhancement in the C-rich tracers at the same azimuth. Late infall of ISM-like material impacting the disk at the observed streamer location naturally satisfies these constraints, and could additionally explain a few dynamical features \citep[i.e., the spirals, the kinematics,][submitted to ApJ]{Calcino2025arXiv251005601C} and therefore provides a compelling explanation for AB~Aur.

\subsubsection{C-rich chemistry on the south side and their link to protoplanets}\label{sec:protoplanet}

While late infall provides a natural explanation for the O-rich SO-bright region, it does not by itself explain why the opposite side of the disk is strongly C$_2$H-bright and SO-faint. The most straightforward way to produce such a C-rich sector is to remove oxygen-bearing volatiles from the local gas phase. In an evolved, massive disk like AB~Aur, the most efficient mechanism to do so is the removal of CO-, CO$_2$- and H$_2$O-rich ices by earlier radial drift \citep{KrijtEtal2020,MahEtal2023}. In this scenario, the outer disk gas, as we observed, would start with a stellar-like abundance and gradually increase its C/O as time evolves. 

On the contrary, if pebble growth and drift are inefficient at the outer disk, e.g., due to the fragile nature of the icy pebbles \citep{MusiolikEtal2016a,MusiolikWurm2019,JiangEtal2024}, the expected high C/O may never be reached by depleting the O-rich ices. A different approach to enhance the gas-phase C/O ratio would be to preferentially sublimate C-rich ices, which have a lower sublimation temperature \citep{OebergEtal2011}. In the cold outer regions of the disk, hydrodynamical simulations indicate that accreting planets embedded in disks generate an azimuthally localized heating source that efficiently sublimes volatiles off the pebbles \citep[e.g.,][]{CleevesEtal2015,JiangEtal2023}. In this scenario, the accretion heating of protoplanet(s) on the southern side of the disk would preferentially release C-rich ices such as CH$_4$ into the vapor phase, driving the local C/O ratio above unity. The enhanced C/O then boosts the C$_2$H abundance, which might reproduce an azimuthal asymmetry on its own if the Keplerian shearing hasn't well mixed the vapor.

The AB~Aur disk exhibits several signatures of ongoing planet formation, including large-scale spirals and substructures in CO gas and scattered light \citep[e.g.,][]{TangEtal2017,BoccalettiEtal2020}, which could originate from potential planets, as demonstrated in dedicated hydrodynamic simulations \citep{DongEtal2016c,FuenteEtal2017}. Additionally, there is a protoplanet candidate exhibiting NIR \citep{CurrieEtal2022} as well as H$\alpha$ emission \citep{CurrieEtal2025}. Notably, all of these features appear on the south side of the disk, see \fg{C2H_planet}. If the southern sector hosts one or more such accreting planets, the carbon-rich ices covered on pebbles formed beyond the CH$_4$ snowline will be intercepted and retained, and boost the gas-phase C/O due to planetary accretion heating. 

\begin{figure}
    \centering
    \includegraphics[width=0.9\linewidth]{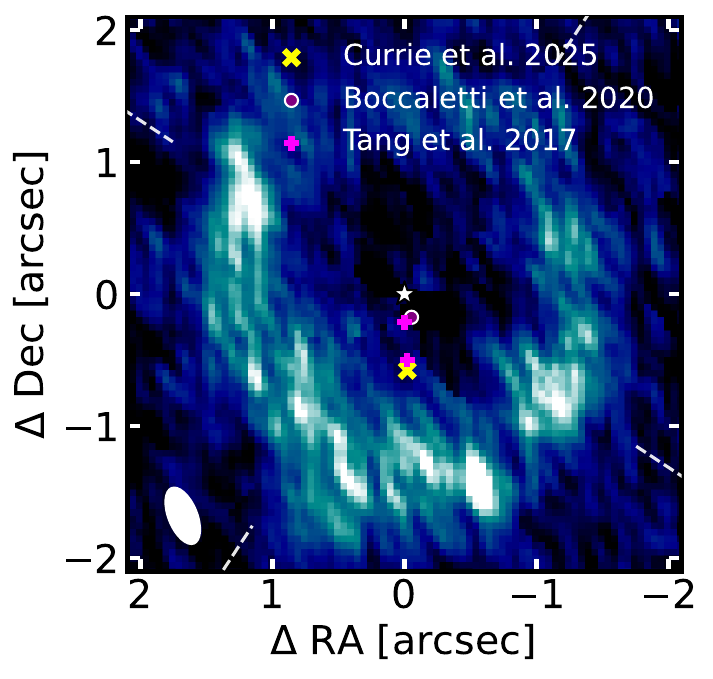}
    \caption{
    The location of the asymmetric C$_2$H ring (four-components staked peak intensity map) relative to the protoplanet candidates from \citet[][pink]{TangEtal2017}, \citet[][purple]{BoccalettiEtal2020}, and \citet[][yellow]{CurrieEtal2022,CurrieEtal2025}. 
    }
    \label{fig:C2H_planet}
\end{figure}

The planet-induced carbon enhancement scenario makes several testable predictions. If the southern sector hosts locally elevated effective C/O, other tracers of UV- and hydrocarbon-rich chemistry, such as CN or cyclic hydrocarbons (e.g., c-C$_3$H$_2$), might also enhance emission in the same azimuthal region. Alternatively, if localized accretion heating is sufficiently strong, it could produce azimuthal asymmetries in higher-excitation molecular tracers, such as the high-energy H$_2$CO transitions recently reported in PDS~70 \citep{RampinelliEtal2024}.

\subsection{Comparison with other Herbig disks}

To place AB~Aur in the broader context of Herbig disk chemistry, we compare its integrated C$_2$H, CS, and SO line fluxes with those measured in other Herbig disks compiled by \citet{BoothEtal2026} (see \fg{comparison_herbig}, where the original line flux are from the Table 2 and 6 of \citet{BoothEtal2026}). All fluxes are scaled to a common distance of 150~pc. Assuming optically thin LTE emission (\se{columndensity}), we further convert the measured line fluxes to the equivalent fluxes of a common set of benchmark transitions for each species. Specifically, we adopt SO $J_N = 6_5-5_4$, CS $J = 5-4$, and C$_2$H $N=3-2,\ J=\tfrac{5}{2}-\tfrac{3}{2}$, corresponding to the transitions analyzed in this work. We assume $T_{\rm rot}=20$~K for CS, and $T_{\rm rot}=40$~K for both SO and C$_2$H, motivated by the rotational diagram analysis presented in \se{rotation}. In addition to the data compiled by \citet{BoothEtal2026}, we include the SO $J_N = 7_7-6_6$ measurement of HD~163296 recently reported by \citet{YamatoEtal2026} and SO $J_N = 6_5-5_4$ measurement of MWC~758 reported by \citet{ZagariaEtal2025}.

Among the disks in the comparison sample, AB~Aur exhibits one of the weakest detected C$_2$H line fluxes despite possessing relatively bright SO emission and a CS flux comparable to those of other Herbig disks.

To quantify the relationships between the molecular line fluxes, we calculate the Pearson correlation coefficient ($r$) using the logarithm of the distance-scaled line fluxes. The corresponding two-sided $p$-value is obtained from the Pearson correlation test under the null hypothesis of no linear correlation. The coefficients of each pair of molecules, as well as the CS/SO ratio versus C$_2$H flux, are listed in \fg{comparison_herbig}. Owing to the limited sample size and the presence of non-detections, we report the correlation coefficients both using only detections (values outside parentheses) and after including upper limits (values in parentheses). For the latter, the upper limits are conservatively treated as measurements at their limiting values, allowing all disks to be included in the comparison while acknowledging that the resulting correlation coefficients should be regarded as qualitative rather than statistically rigorous. Consequently, the reported $p$-values for the censored sample should be interpreted with caution, and the correlations are used only to illustrate general trends in the data.

Although the sample remains small and several measurements are upper limits, our comparison is consistent with the positive correlation between CS and C$_2$H reported by \citet{BoothEtal2026}. We also find tentative evidence that disks with larger CS/SO ratios tend to exhibit stronger C$_2$H emission, consistent with the expectation that both C$_2$H and the CS/SO ratio are sensitive to the gas-phase C/O ratio \citep[e.g.,][]{LeGalEtal2021,BosmanEtal2021b}.

The molecular inventory of AB~Aur therefore might differ from chemically carbon-rich Herbig disks such as HD~163296 and MWC~480, which exhibit much brighter C$_2$H emission and have been interpreted as possessing gas-phase C/O ratios exceeding $\sim1.5$ \citep{BosmanEtal2021b}. Instead, this may partly reflect the ongoing infall onto AB~Aur may continuously replenish or mix material with a more ISM-like C/O ratio into the outer disk. Such replenishment would counteract the chemical evolution driven by dust settling and radial drift, inhibiting the development of the C$_2$H-rich chemistry observed in more evolved Herbig disks while remaining consistent with the relatively strong SO emission observed in AB~Aur.

\begin{figure*}
    \centering
    \includegraphics[width=0.99\linewidth]{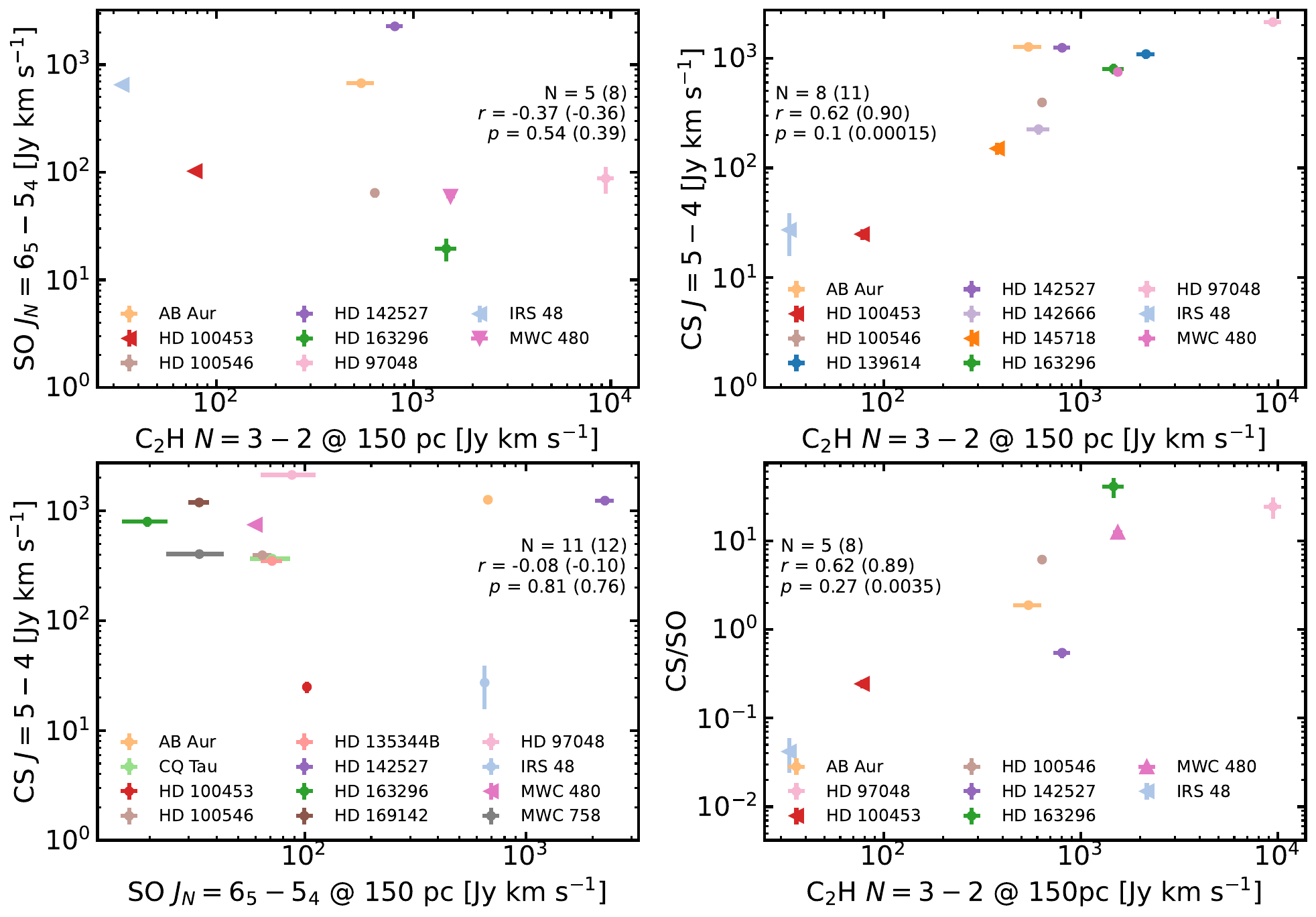}
    \caption{Comparison of the integrated C$_2$H, CS, and SO line fluxes of AB~Aur with those of other Herbig disks compiled by \citet{BoothEtal2026}. All line fluxes are scaled to a common distance of 150~pc and converted to the benchmark transitions analyzed in this work under the assumptions of optically thin LTE emission (see main text). Colored symbols identify individual disks, while arrows denote $3\sigma$ upper limits. The number of detections used in the Pearson correlation analysis is indicated as $N$, with the total sample size including upper limits given in parentheses. Likewise, the Pearson correlation coefficient ($r$) and corresponding two-sided $p$-value are reported for detections only (outside parentheses) and for the full sample with upper limits conservatively treated as measurements at their limiting values (inside parentheses).}
    \label{fig:comparison_herbig}
\end{figure*}

\subsection{Carbon isotopic ratio based on HCO$^+$ and H$^{13}$CO$^+$}

\begin{figure}
    \centering
    \includegraphics[width=0.98\linewidth]{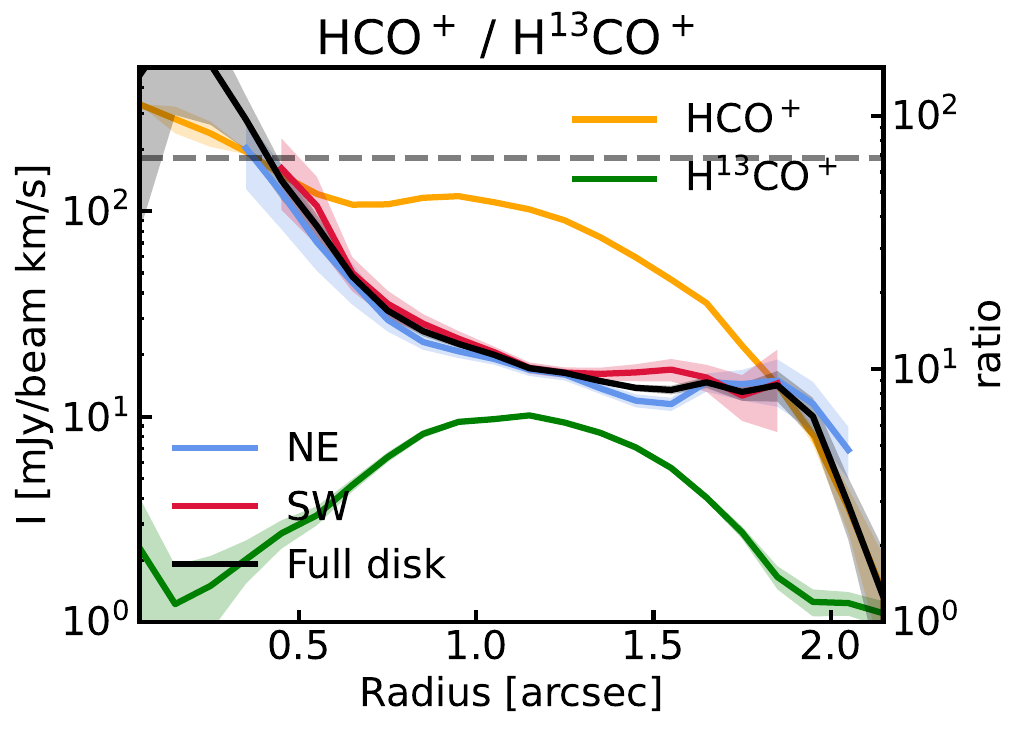}
    \caption{Integrated intensity radial profiles of HCO$^+$/H$^{13}$CO$^+$ and their corresponding ratio. Black lines show azimuthally averaged values. Blue and red lines are the profiles along blue- and red-shift sides of the disk, which show no difference within the error range, despite the azimuthal asymmetries. The horizontal dashed line in the upper panel indicates the local ISM $^{12}$C/$^{13}$C abundance ratio. The HCO$^+$/H$^{13}$CO$^+$ line ratio decreases from values close to the canonical isotopic ratio in the inner disk to only $\sim10$--15 within the ring.}
    \label{fig:HCO_ratio}
\end{figure}

The radial profile of the HCO$^+$/H$^{13}$CO$^+$ integrated intensity ratio provides a diagnostic of the carbon isotopic chemistry and the optical depth of the ionized molecular gas. As shown in Figure~\ref{fig:HCO_ratio}, the ratio is comparable to the local interstellar $^{12}$C/$^{13}$C isotopic abundance ratio ($\sim69$) in the innermost disk, but decreases rapidly with radius. At the location of the prominent HCO$^+$(H$^{13}$CO$^+$) as well as continuum ring ($\sim1''$--$2''$), the observed HCO$^+$/H$^{13}$CO$^+$ integrated intensity ratio is only $\sim10$--15, substantially lower than the canonical isotopic ratio.

A reduced isotopologue line ratio may, in principle, result from isotope-selective chemistry \citep[e.g.,][]{BerginEtal2024a}. However, the coincidence of the low ratio with the bright HCO$^+$ emission ring strongly suggests that optical depth dominates the observed behavior. Assuming that the intrinsic $^{12}$C/$^{13}$C abundance ratio remains equal to the local ISM value of 69, and H$^{13}$CO$^+$ share the same emitting layer as HCO$^+$, the observed line ratio can be expressed as

\begin{equation}
R \equiv 
\frac{1-e^{-\tau_{\rm HCO^{+}}}}
{1-e^{-\tau_{\rm H^{13}CO^{+}}}}
=
\frac{1-e^{-\tau_{\rm HCO^{+}}}}
{1-e^{-\tau_{\rm HCO^{+}}/69}},
\end{equation}
where $\tau_{\rm HCO^{+}} \equiv 69 \tau_{\rm H^{13}CO^{+}}$ are the optical depth of HCO$^+$ and H$^{13}$CO$^{+}$ lines. Inverting this relation using the measured line ratios yields $\tau_{\rm HCO^+}\approx6$--11 over the radial range of $1''$--$2''$. This indicates that the HCO$^+$ $J=3$--2 emission could be highly optically thick throughout the molecular ring, whereas the H$^{13}$CO$^+$ emission is expected to remain largely optically thin.

The large optical depth has important consequences for interpreting the HCO$^+$ morphology. In the optically thick regime, the observed HCO$^+$ intensity is primarily determined by the excitation temperature and the location of the $\tau\sim1$ surface, rather than being directly proportional to the HCO$^+$ column density. Consequently, the bright HCO$^+$ asymmetry coincident with the continuum blob does not necessarily imply a corresponding enhancement in molecular abundance or gas surface density, but is more naturally explained by a local temperature enhancement. This interpretation is consistent with the C$^{17}$O hyperfine structure analysis of \citet{DutreyEtal2024}, who likewise inferred higher gas temperatures at the location of the continuum blob than elsewhere along the continuum ring.

Nevertheless, the relatively high HCO$^+$/H$^{13}$CO$^+$ ratio observed inside the cavity ($\lesssim70$ au) may also reflect isotope-selective photochemistry. In the tenuous inner disk, the stronger stellar FUV radiation and lower gas column density reduce dust and molecular shielding, allowing selective photodissociation of the less abundant $^{13}$CO isotopologue to become more efficient than for $^{12}$CO. Because CO is the dominant progenitor of HCO$^+$ through ion–molecule chemistry, an enhanced $^{12}$CO/$^{13}$CO abundance ratio would naturally propagate into an elevated HCO$^+$/H$^{13}$CO$^+$ abundance ratio. Disk chemical models predict that selective photodissociation can increase the $^{12}$CO/$^{13}$CO ratio above the elemental abundance ratio in UV-irradiated surface layers, particularly once the gas column density becomes sufficiently low \citep[e.g.,][]{VisserEtal2009}. Therefore, while we adopt the canonical ISM $^{12}$C/$^{13}$C ratio when estimating the HCO$^+$ optical depth in the molecular ring, this assumption may not strictly hold in the innermost disk.

\subsection{Implication for giant planet atmosphere}

Taken together, our results indicate that AB~Aur hosts both external (infall–driven) and internal (planet–driven) mechanisms that comparably shape the disk thermal and chemical structures. In the north, late infall locally resets the gas to an O–rich state and enhances SO; in the south, ongoing planet formation maintains a C–rich gas that is bright in C$_2$H. The azimuthal C/O gradient might be a direct signature of the interplay between environmental accretion and planet–induced substructures, implying that the chemistry sampled by forming planets depends sensitively on their location relative to infall streams and pressure traps.

If infall is intermittent, with multiple streamer impacts over the disk lifetime \citep{WinterEtal2024,PadoanEtal2025}, recurrent episodes can repeatedly drive the affected sectors toward an ISM–like, O–rich composition, counteracting the secular tendency of disk evolution to raise C/O via solid–gas decoupling \citep{BosmanEtal2021a,CalahanEtal2023,JiangEtal2023}. This “chemical reset” should be most effective at late stages, when the disk’s intrinsic gas reservoir is smaller and therefore more susceptible to pollution by external material.

This framework offers a potential link to observations of sub–stellar companions. The majority of wide–separation objects show near–solar atmospheric C/O \citep{HochEtal2023,XuanEtal2024,BerginEtal2024b}, even though planet–forming gaps in disks are often inferred to be C–rich \citep{BosmanEtal2021b}. Episodic infall provides a natural source of O–rich gas that may potentially dilute and reset the local C/O in the birth environment, reconciling companion atmospheres with the chemically heterogeneous conditions expected in structured, dynamically evolving disks. 

Recent atmospheric studies have also begun to measure carbon isotopic ratios in directly imaged giant planets \citep[e.g.,][]{ZhangEtal2024c}, providing an additional tracer of planet formation and disk chemical evolution. If selective photodissociation enhances the gas-phase $^{12}$C/$^{13}$C ratio in the warm inner regions of Herbig disks, as suggested by the high HCO$^+$/H$^{13}$CO$^+$ ratios measured toward the inner disk, then directly imaged giant planets exhibiting sub-ISM atmospheric $^{12}$C/$^{13}$C ratios are unlikely to have accreted the bulk of their atmospheres from these inner regions. Instead, such isotopic compositions would favor formation in the colder outer disk, where low-temperature isotope chemistry is expected to dominate. This supports the hypothesis that isotopic fractionation, in addition to elemental abundances, may encode information on the natal disk environment and planet formation history \citep[e.g.,][]{BerginEtal2024a}.

We emphasize that these implications remain model-dependent, but the observed chemical segregation demonstrates that protoplanetary disks need not be azimuthally homogeneous in their elemental composition.

\section{Summary}\label{sec:summary}

We present NOEMA 1.2\,mm line observations of the Herbig Ae disk AB~Aur and investigate the origin of its azimuthal chemical structure.

\begin{itemize}

\item Molecular line imaging reveals a pronounced azimuthal anti-correlation between SO and C$_2$H at radii of $\sim200$–220\,au, slightly beyond the continuum ring location of $\sim160$\,au. All detected SO transitions peak in the northern sector, spatially coincident with the inferred streamer--disk interaction region, whereas C$_2$H peaks in the southern sector on the opposite side of the disk. In contrast, CS~$J{=}5{-}4$ forms a nearly axisymmetric ring at the same radii.

\item HCN and HCO$^+$ exhibit a different morphology: both peak exactly near the dust continuum overdensity along the main ring, and HCO$^+$ additionally shows spatially resolved emission inside the dust cavity. H$^{13}$CO$^+$ follows the overall ring structure. Their behavior contrasts with the SO and C$_2$H anti-correlations and further indicates that the azimuthal structure is chemically selective rather than driven solely by variations in gas surface density.

\item Multi-transition rotational-diagram analyses of SO and CS show that the SO-bright northern sector exhibits both elevated rotational temperatures and enhanced SO column densities, whereas CS is nearly symmetric in both temperature and column density despite tracing similar radii. The substantially higher rotational temperatures inferred for SO suggest that SO and CS originate from different disk layers and/or chemical components. Although excitation contributes to the observed intensity contrasts, it cannot simultaneously reproduce the SO enhancement, the C$_2$H suppression in the north, and the symmetric CS ring.

\item Comparison with gas-grain chemical models suggests that the observed SO and C$_2$H anti-correlation is consistent with azimuthal variations in effective C/O. The SO-bright northern sector is most consistent with relatively oxygen-rich conditions, while the C$_2$H-bright southern sector favors hydrocarbon-rich chemistry and comparatively elevated C/O. These inferences remain model-dependent but are supported by the behavior of the CS/SO ratio \citep[e.g.,][]{LeGalEtal2021,KeyteEtal2023,HuangEtal2024} and by the sensitivity of C$_2$H abundance to C/O \citep[e.g.,][]{BerginEtal2016,BosmanEtal2021b}.

\item We propose two non-exclusive scenarios. In the north, late infall associated with the streamer may induce localized heating in the surface layer and release O-bearing ices, enhancing SO through oxygen-rich chemistry (\Se{infall}). In the south, disk substructure and potential planet formation may modify the local volatile balance and UV environment, promoting hydrocarbon-rich chemistry and enhanced C$_2$H emission (\Se{protoplanet}).

\item These scenarios yield testable predictions. Shock and infall tracers (e.g., warm H$_2$CO, CH$_3$OH, SO$_2$) and subtle inflow kinematics should be enhanced in the northern sector, while high-C/O tracers (e.g., CN, C$_3$H$_2$) and signatures of embedded accreting planets should coincide with the southern C$_2$H-bright region.

\item By comparing the integrated molecular line fluxes of AB~Aur with those of other Herbig disks, we find that AB~Aur exhibits comparably weak C$_2$H emissions despite relatively bright SO emission and typical CS emission. This behavior contrasts with chemically carbon-rich Herbig disks such as HD~163296, MWC~480 and HD~97048, who host very bright C$_2$H emissions while have weak or non-detection of SO. This is consistent with our hypothesis that ongoing late infall replenishes the outer disk with oxygen-rich material, suppressing the development of extreme hydrocarbon-rich chemistry.

\item The HCO$^+$/H$^{13}$CO$^+$ line ratio indicates that HCO$^+$ $J=3$--2 is strongly optically thick ($\tau\sim6$--11) across the molecular ring, whereas H$^{13}$CO$^+$ remains largely optically thin. The elevated HCO$^+$/H$^{13}$CO$^+$ ratio in the inner cavity also points to an enhanced gas-phase $^{12}$C/$^{13}$C ratio, implying that giant planets with canonical or sub-ISM atmospheric $^{12}$C/$^{13}$C ratios are unlikely to have accreted the bulk of their atmospheres from the inner disk. These results imply that the HCO$^+$ asymmetry is primarily sensitive to local excitation temperature rather than directly tracing gas column density, and further support the assumption that the much fainter CS, SO, and C$_2$H transitions are at least marginally optically thin.

\item The results demonstrate that late infall and planet formation can generate significant azimuthal chemical heterogeneity within a single disk. Such spatial variation in effective C/O may influence the compositions of forming planets and provide a potential framework for interpreting atmospheric abundance measurements of wide-separation companions.

\end{itemize}

The azimuthal chemical differentiation in AB~Aur suggests that dynamically active protoplanetary disks could be chemically inhomogeneous at a given radius. In disks undergoing late infall, substructure formation, or ongoing planet formation, localized variations in temperature, irradiation, and volatile partitioning may lead to spatial variations in effective C/O and molecular abundances. Such heterogeneity has important implications for interpreting disk-integrated chemical diagnostics and for understanding the origin of compositional diversity in forming planets. 

Determining whether this behavior is common requires high-angular-resolution chemical inventory surveys across a statistically meaningful sample of structured disks. Systematic multi-line observations targeting complementary tracers of C-, O-, and S-bearing species will be essential for establishing how disk dynamics, volatile transport, and planet formation jointly regulate azimuthal chemical structure.

\begin{acknowledgements}
    We thank the anonymous referee for their thoughtful and constructive comments.
    H.J. thanks Dr.~Alice Booth for sharing the CS $J = 7$-$6$ cube reduced by Dr.~Milou Temmink.
    This project has received funding from the European Research Council (ERC) under the European Union’s Horizon 2020 research and innovation programme (PROTOPLANETS, grant agreement No. 101002188). Views and opinions expressed are, however, those of the author(s) only and do not necessarily reflect those of the European Union or the European Research Council Executive Agency. Neither the European Union nor the granting authority can be held responsible for them. 
    D.S. is funded by the Deutsche Forschungsgemeinschaft (DFG, German Research Foundation) – project number: 550639632. 
    P.R.M. is a member of project PID2022-137980NB-I00, funded by MCIN/AEI/10.13039/501100011033/FEDER UE.
    This work is based on observations carried out under project number W24BR with the IRAM NOEMA Interferometer. IRAM is supported by INSU/CNRS (France), MPG (Germany) and IGN (Spain).
    This paper makes use of the following ALMA data: 2012.1.00303.S, 2021.1.00690.S. ALMA is a partnership of ESO (representing its member states), NSF (USA) and NINS (Japan), together with NRC (Canada), NSTC and ASIAA (Taiwan), and KASI (Republic of Korea), in cooperation with the Republic of Chile. The Joint ALMA Observatory is operated by ESO, AUI/NRAO and NAOJ. 
    The authors acknowledge the use of Grammarly to correct grammatical errors and improve the writing style of this manuscript. This work makes use of the \texttt{IMAGER} and \texttt{GILDAS} software to reduce and analyze the data (See https://imager.oasu.u-bordeaux.fr and http://www.iram.fr/IRAMFR/GILDAS).
\end{acknowledgements}
\bibliographystyle{aa}
\bibliography{ads}

\begin{appendix}

\section{Continuum imaging and gallery}\label{app:continuum}

We extracted the 1.2\,mm continuum emission from the line-free channels of the PolyFiX setup in all four sidebands. After excluding spectral regions with strong molecular emission, the remaining channels were averaged within each sideband to maximize the signal-to-noise ratio. The continuum visibilities were self-calibrated following the procedure described in \se{observation} and imaged in CASA using the \texttt{multiscale} deconvolver.

To assess the robustness of the continuum morphology against imaging choices, we produced images at four representative frequencies corresponding to the centers of the four PolyFiX sidebands. For each frequency, we generated maps using four different Briggs robust parameters ($-1.5$, $-0.5$, $+0.5$, and $+1.5$), spanning the range from higher angular resolution to higher sensitivity.

The resulting continuum gallery is shown in \fg{cont_AB_Aur}. Across frequencies and weighting schemes, the continuum emission consistently exhibits a bright, structured ring with a pronounced overdensity in the west sector, and a central cavity, within which an unresolved point source is detected at the stellar position.

\begin{figure*}
    \centering\includegraphics[width=.88\linewidth]{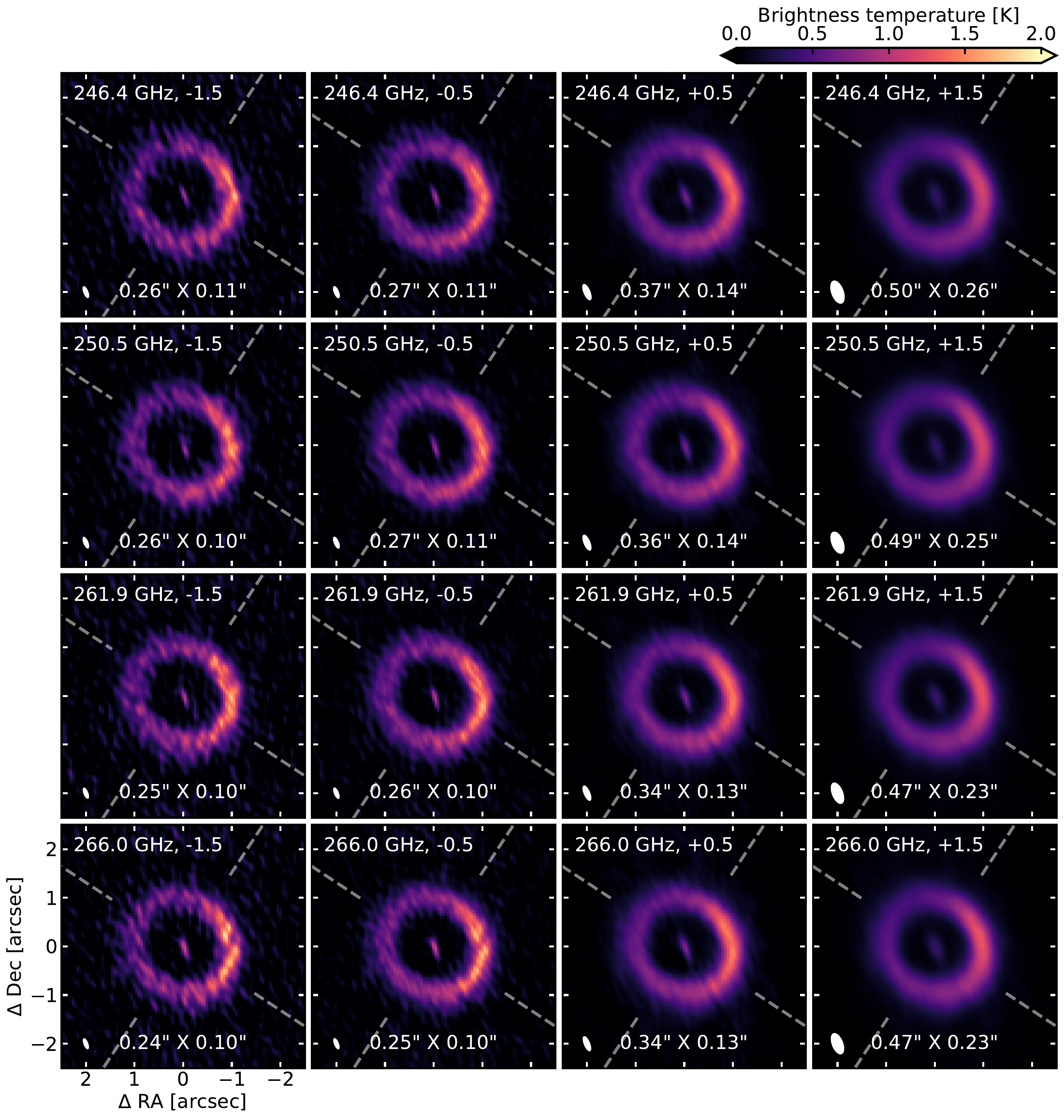}
    \caption{\label{fig:cont_AB_Aur} 
    {\bf Continuum gallery}. From top to bottom: four representative frequencies centered on the four NOEMA PolyFiX sidebands. From left to right: images produced with Briggs robust parameters of $-1.5$, $-0.5$, $+0.5$, and $+1.5$. The frequency and robust parameter are labeled in the upper-left corner of each panel; the synthesized beam is shown in the lower-right corner. Dashed lines indicate the disk major and minor axes.}
\end{figure*}

\section{Channel maps}\label{app:channel_maps}

\begin{figure*}
    \centering\includegraphics[width=.82\linewidth]{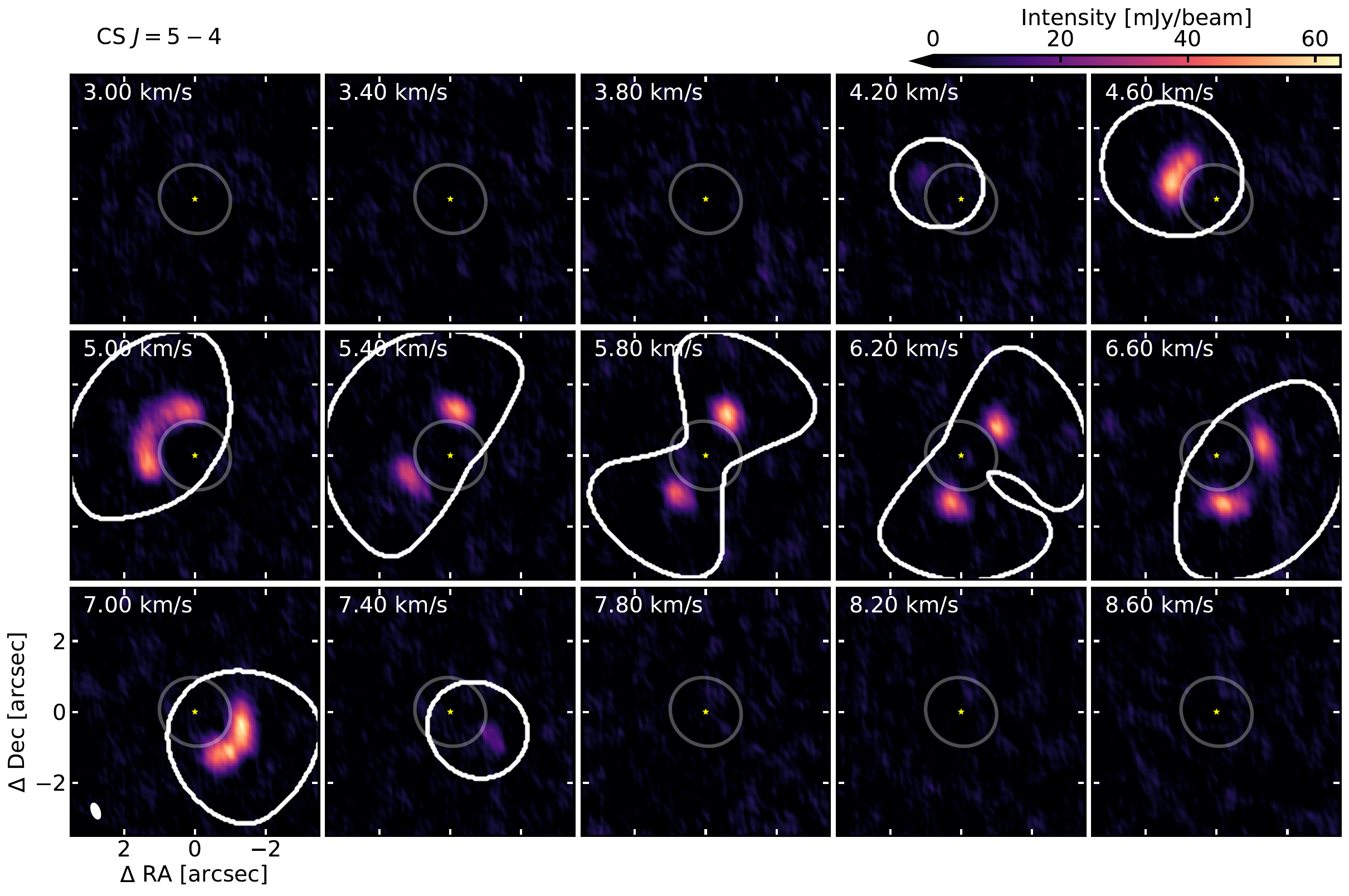}
    \caption{\label{fig:CS_J_N_5-4_chan} Channel maps of CS $J=5-4$.}
\end{figure*}

\begin{figure*}
    \centering\includegraphics[width=.82\linewidth]{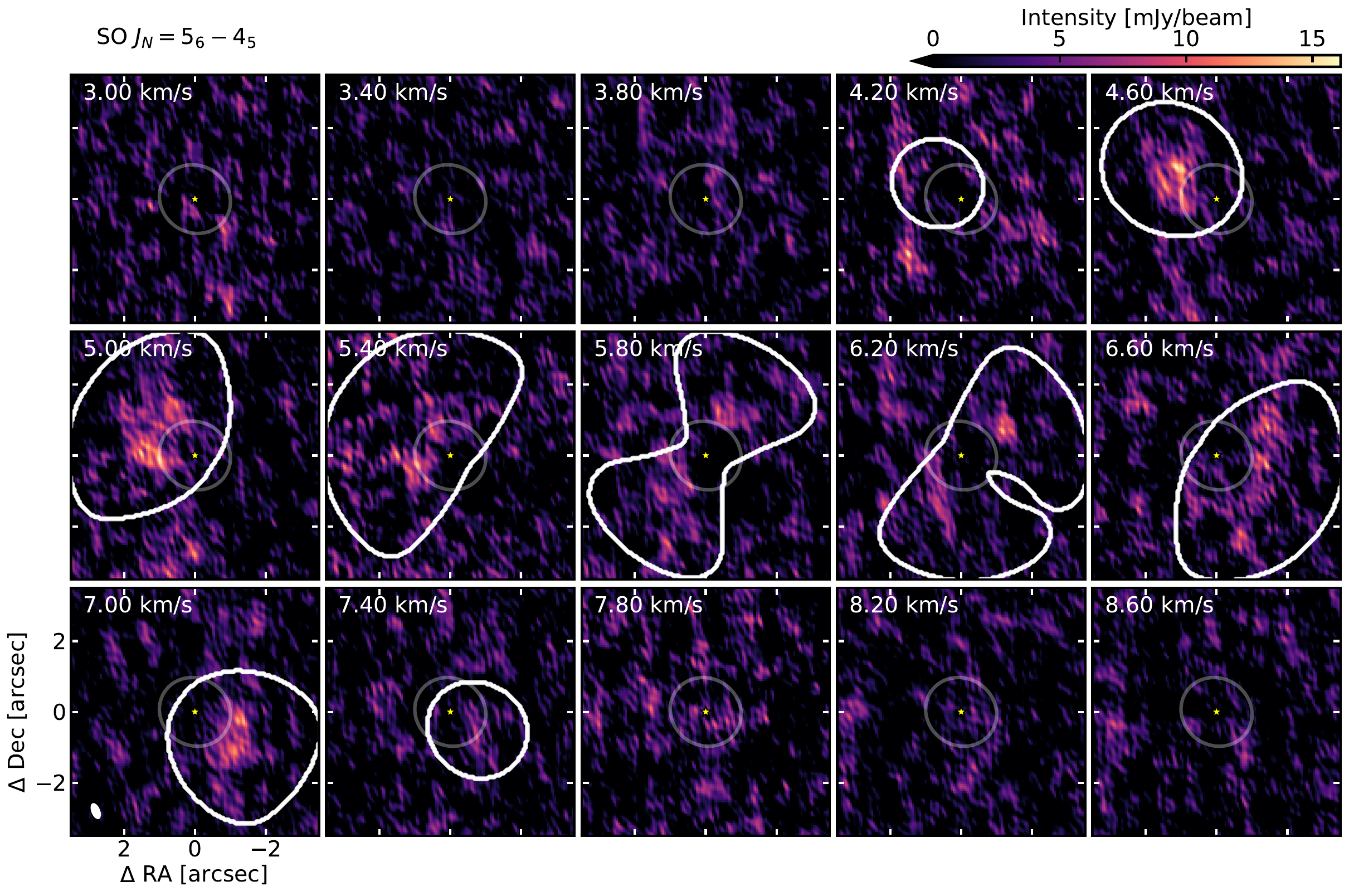}
    \caption{\label{fig:SO_J_N_5_6-4_5_chan} Channel maps of SO $J_N=5_6-4_5$.}
\end{figure*}

\begin{figure*}
    \centering\includegraphics[width=.82\linewidth]{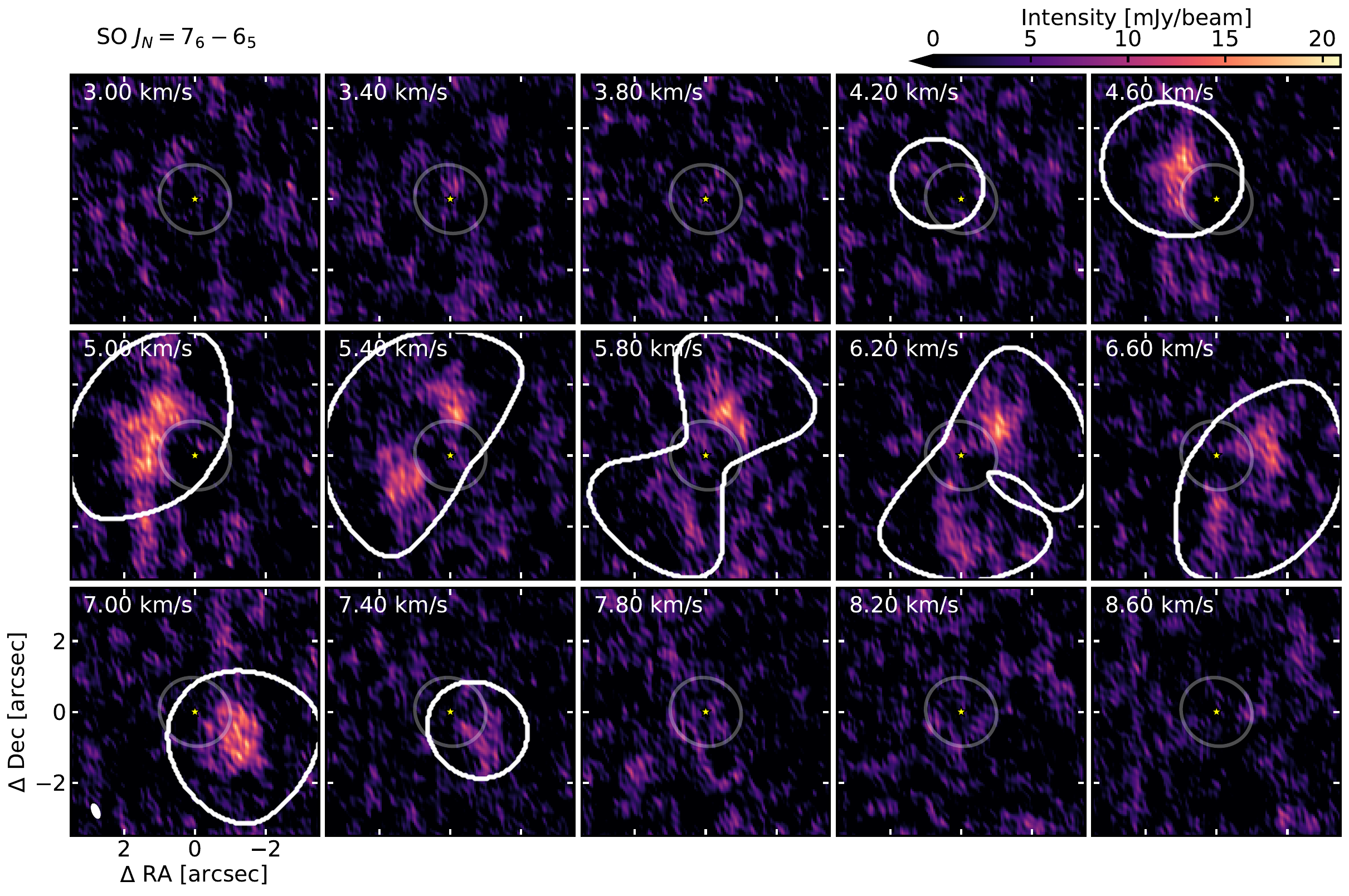}
    \caption{\label{fig:SO_J_N_7_6-6_5_chan} Channel maps of SO $J_N=7_6-6_5$.}
\end{figure*}

\begin{figure*}
    \centering\includegraphics[width=.82\linewidth]{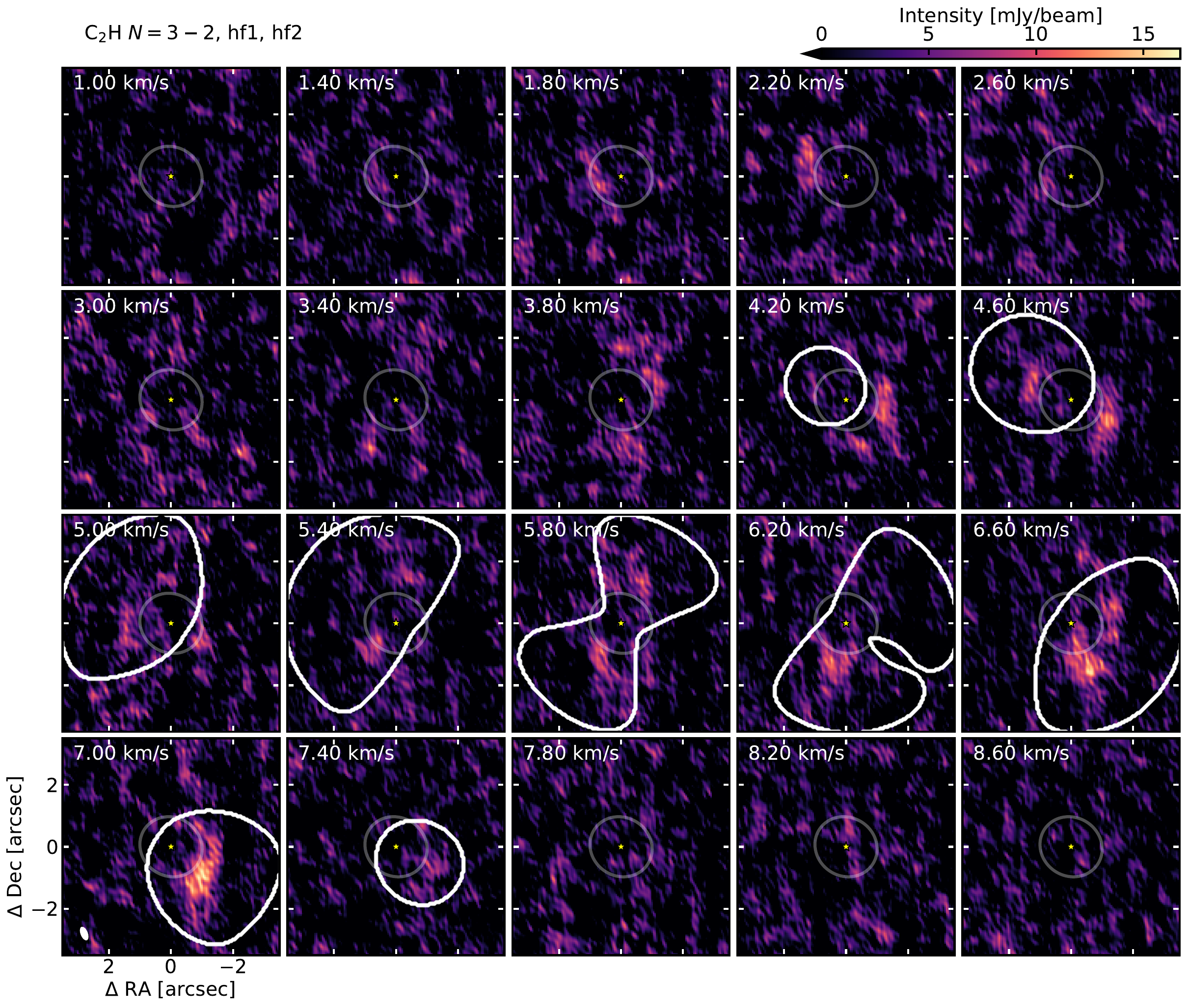}
    \caption{\label{fig:C2H_N_3-2_hf12_chan} Channel maps of C$_2$H $N=3-2,\ J=\tfrac{7}{2}-\tfrac{5}{2},\ F=4-3$ (hf1 and hf2).}
\end{figure*}

\begin{figure*}
    \centering\includegraphics[width=.82\linewidth]{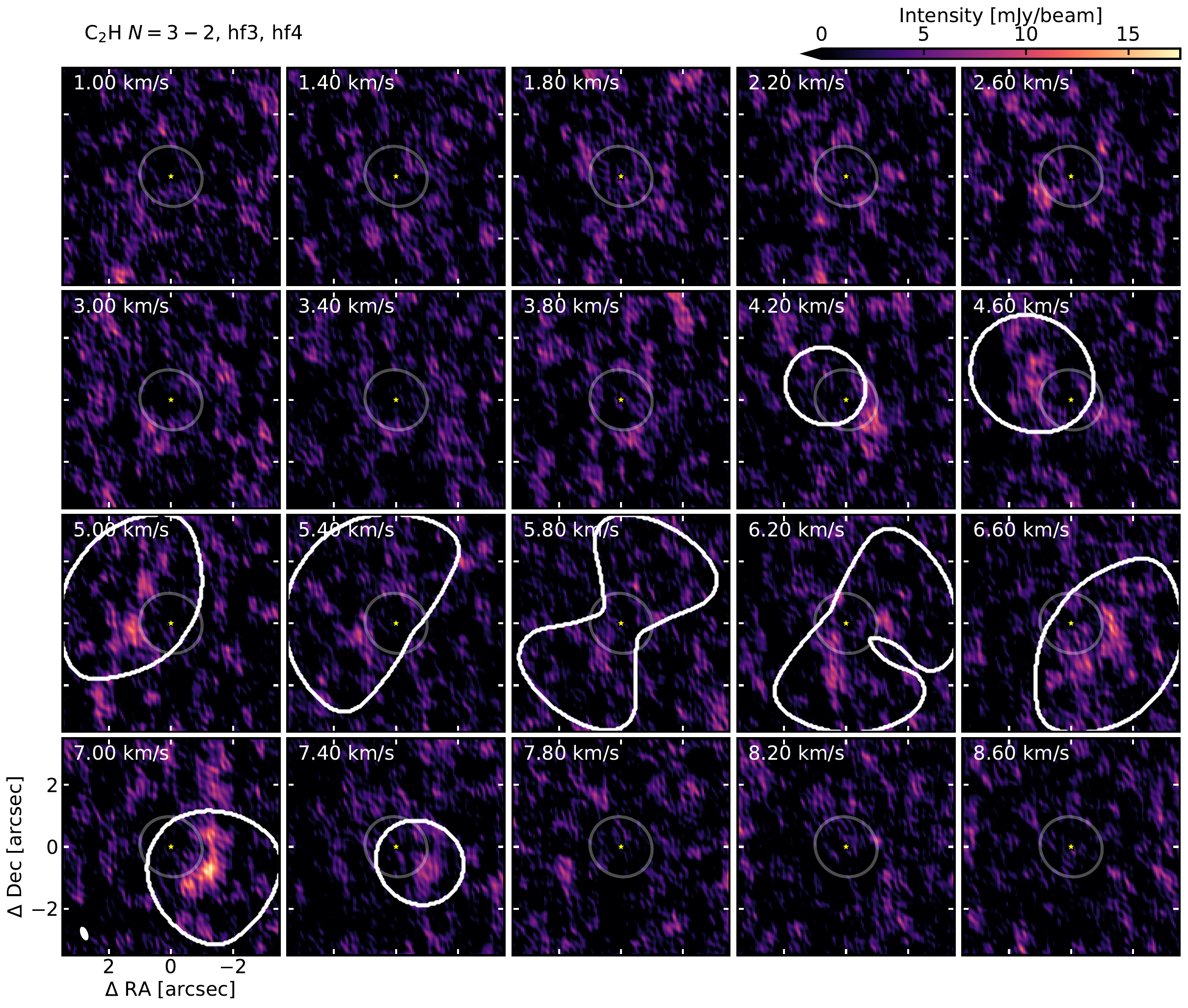}
    \caption{\label{fig:C2H_N_3-2_hf34_chan} Channel maps of C$_2$H $N=3-2,\ J=\tfrac{5}{2}-\tfrac{3}{2},\ F=3-2$ (hf3 and hf4).}
\end{figure*}

\begin{figure*}
    \centering\includegraphics[width=.82\linewidth]{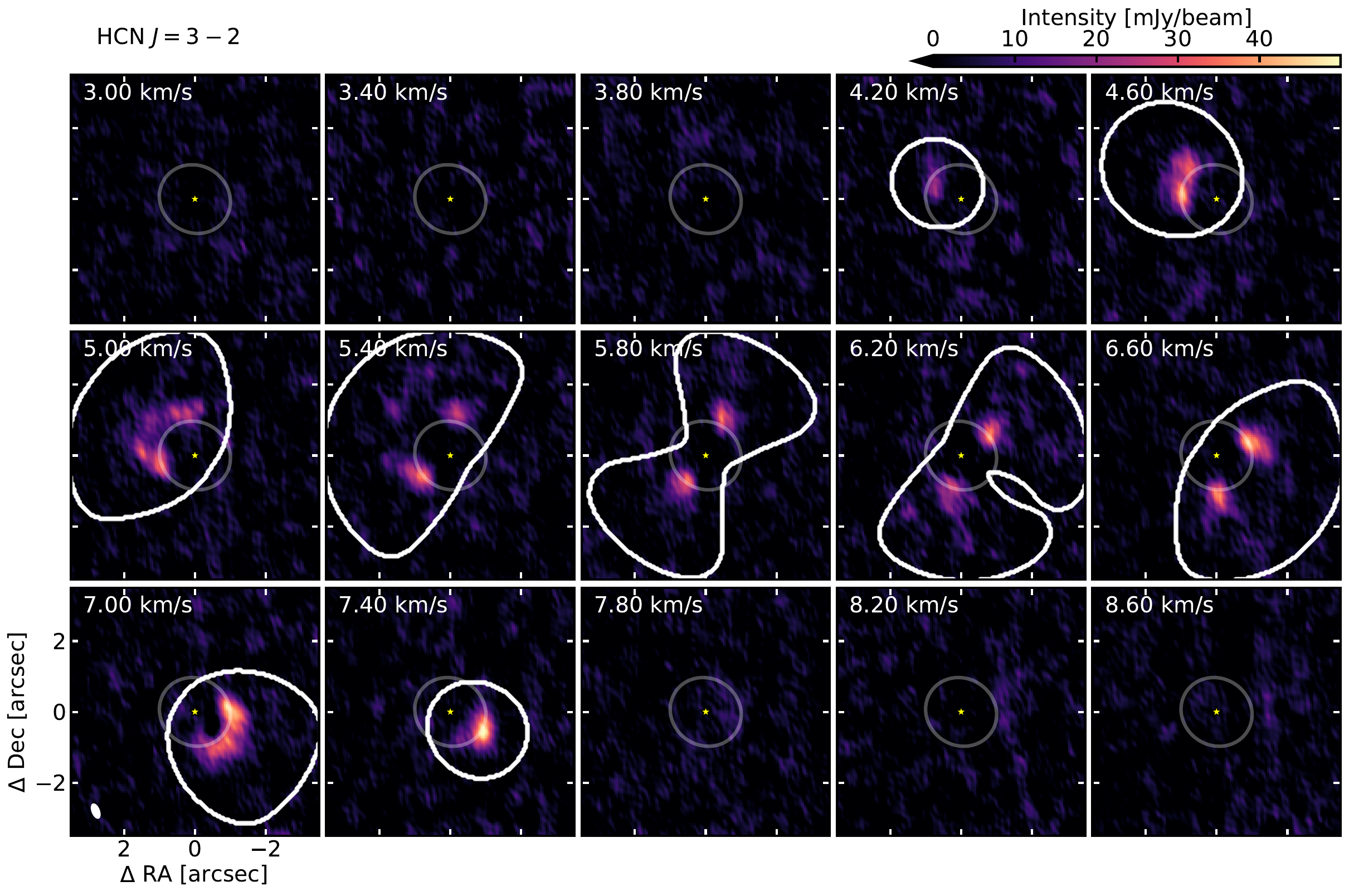}
    \caption{\label{fig:HCN_J_3-2_chan} Channel maps of HCN $J=3-2$.}
\end{figure*}

\begin{figure*}
    \centering\includegraphics[width=.82\linewidth]{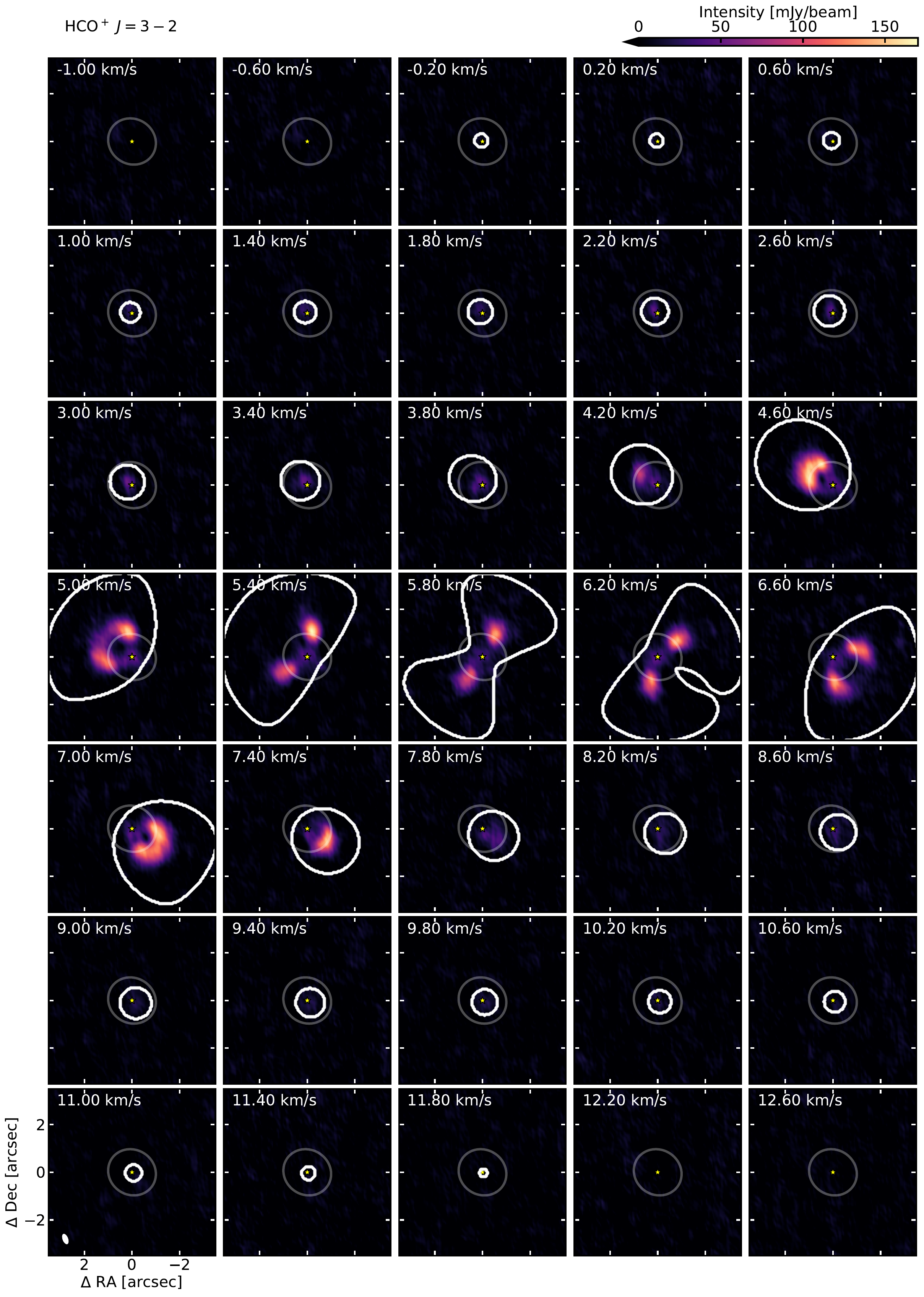}
    \caption{\label{fig:HCO+_J_3-2_chan} Channel maps of HCO$^+$ $J=3-2$.}
\end{figure*}

\begin{figure*}
    \centering\includegraphics[width=.82\linewidth]{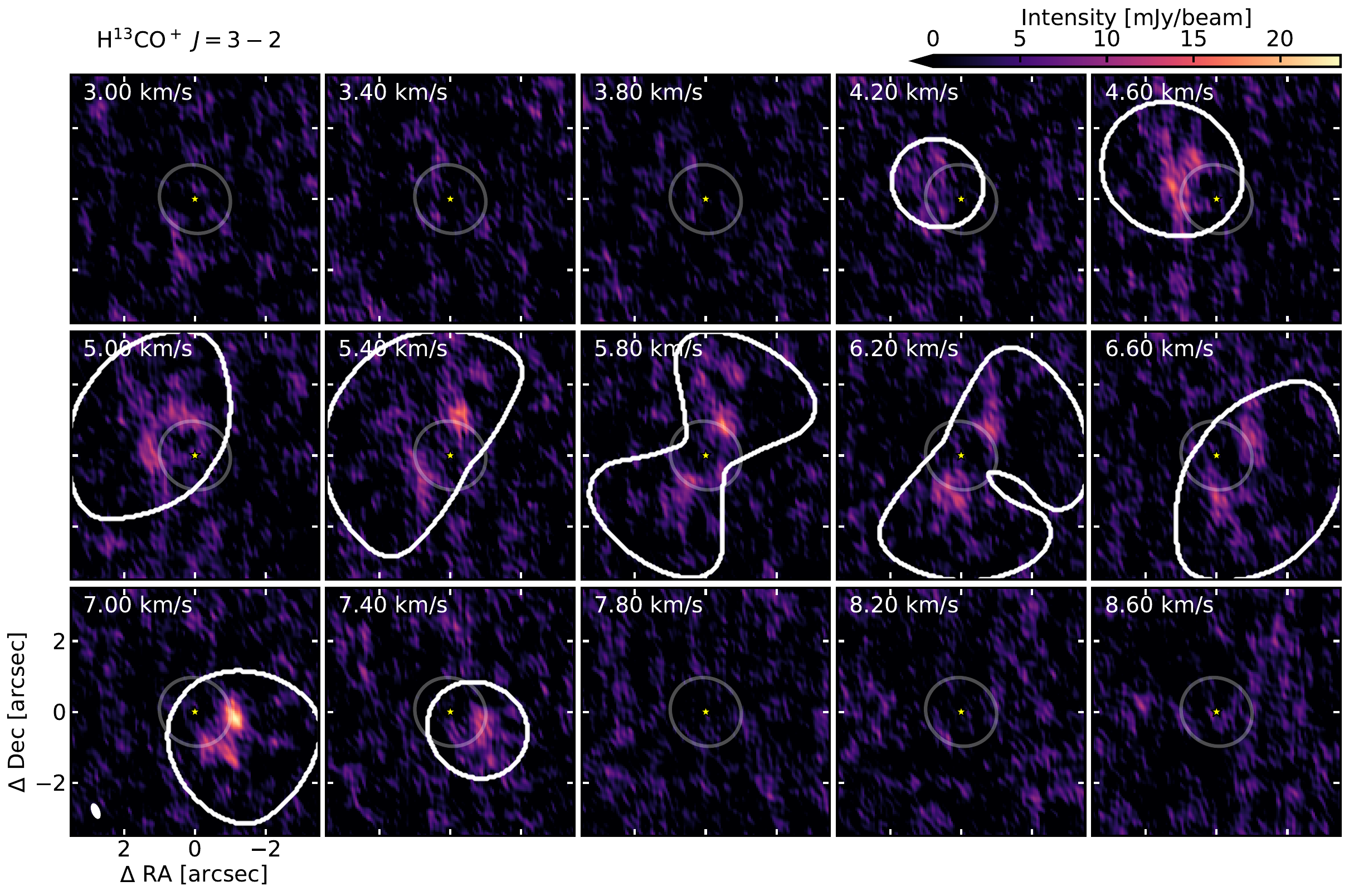}
    \caption{\label{fig:H13CO+_J_3-2_chan} Channel maps of H$^{13}$CO$^+$ $J=3-2$.}
\end{figure*}

We present channel maps of all the detected molecular transitions in Figs.~\ref{fig:CS_J_N_5-4_chan}–\ref{fig:H13CO+_J_3-2_chan}. For consistency, all maps show velocity channels between 3.0 and 8.6\,km\,s$^{-1}$ with a uniform spacing of 0.4\,km\,s$^{-1}$, expect for C$_2$H and HCO$^+$. 

For C$_2$H, as the hf2 and hf4 transitions are very close to the brighter hf1 and hf3 (see the teardrop plots in \fg{AB_Aur_profiles}), we show them together with hf1 and hf3 by extending the lower end of the velocity ranges to 1.0 \,km\,s$^{-1}$. The zero velocity is centered at the rest frequencies of hf1 and hf3. The systematic velocities for hf2 and hf4 are therefore $\sim$3.4\,km\,s$^{-1}$ and $\sim$3.0\,km\,s$^{-1}$, respectively, in \fg{C2H_N_3-2_hf12_chan} and \fg{C2H_N_3-2_hf34_chan}.

For HCO$^+$, since the inner disk is clearly detected, we present velocity channels ranging from -1.0 to 12.6\,km\,s$^{-1}$ to include the faint inner disk emission, which deviates further from the systematic velocity. 

In each panel, the stellar position is marked by a yellow star and the continuum ring peak radius is indicated by a gray circle. The white counter in each panel shows the Keplerian mask used to generate the peak and integrated maps presented in the main text. These maps show the spatial distribution and kinematic structure of the emission across the disk.

\section{Inner disk of HCO$^+$}\label{app:hcop_inner_disk}

From the moment zero and peak intensity maps, we can see that there is an offset between the central emission of HCO$^+$ and that of the continuum. In \fg{HCOp_inner_disk}, we integrated the robust = 0.5 cube over the three different velocity ranges to demonstrate that this offset is caused by asymmetry in the inner disk. 

The blue-shifted component, integrated from -0.2 km/s to 3.8 km/s, is approximately twice as bright as the red-shifted component, integrated from 7.8 km/s to 11.8 km/s. Given the spatial and spectral resolution, it is difficult to determine the origin of this asymmetry in the inner disk. However, we note that \citet[][submitted to AJ]{Blakely2026arXiv260207731B} report a significant detection of an anomaly caused by a companion in the cavity of AB~Aur, combining Gaia and Hipparcos astrometry. If the potential companion is within the sub-stellar regime, its location would be consistent with that of the protoplanet candidate (see \se{protoplanet} and \fg{HCOp_planet}). Yet, if the anomaly is instead caused by a low-mass star instead, it would be located close to the brighter, blue-shifted inner disk. 

\begin{figure*}
    \centering
    \includegraphics[width=.9\linewidth]{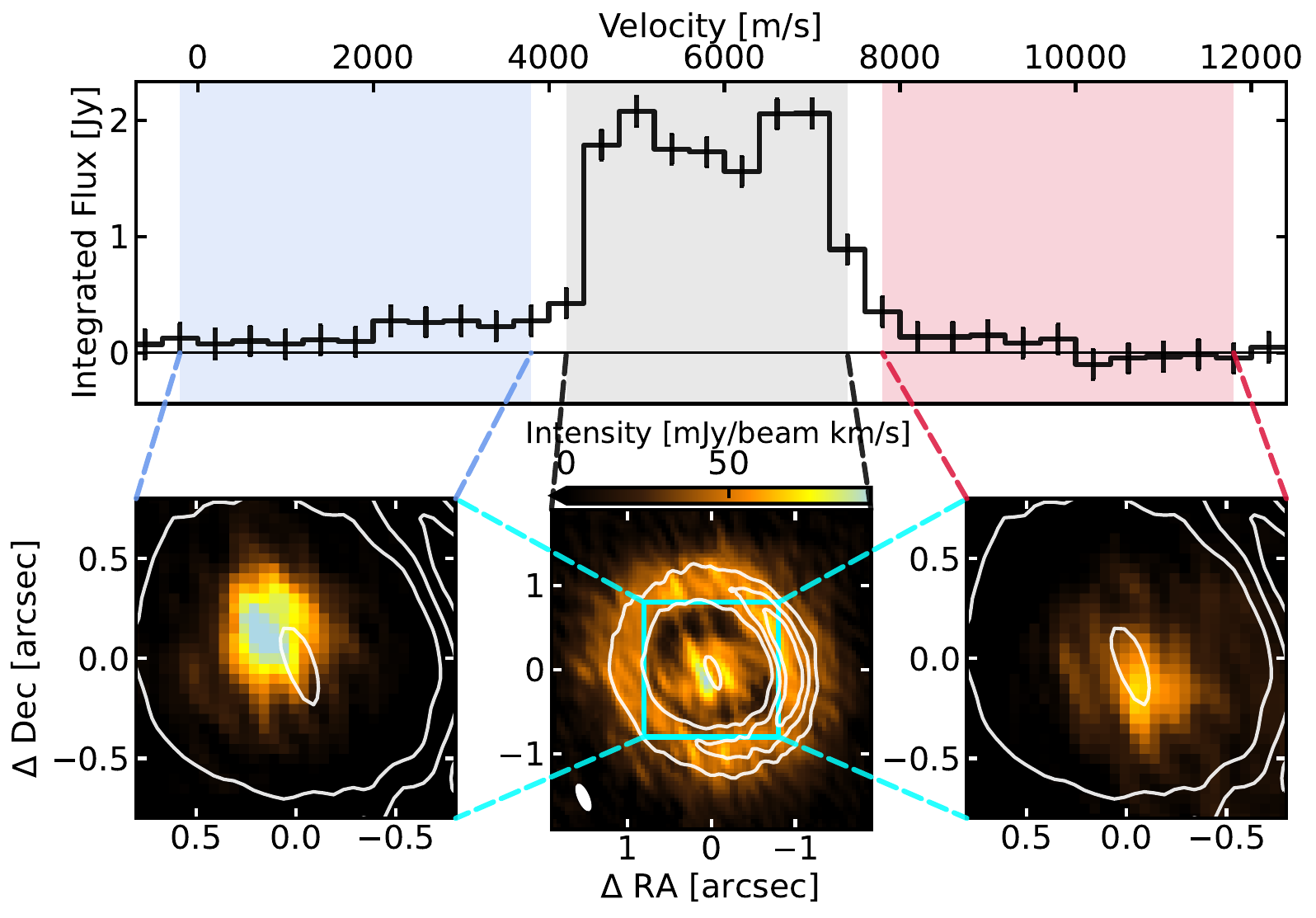}
    \caption{The spatial and kinematic structure of the HCO$^+$ emission. A cube imaged with a robust number of 0.5 is used. The top panel shows the integrated spectrum, with the velocity ranges used for the moment maps highlighted as blue (blue-shifted), gray (systemic), and red (red-shifted) intervals. The bottom row presents the corresponding moment-0 maps without any masking. The blue-shifted inner disk is significantly brighter than the red-shifted part. }
    \label{fig:HCOp_inner_disk}
\end{figure*}

\begin{figure*}
    \centering
    \includegraphics[width=0.5\linewidth]{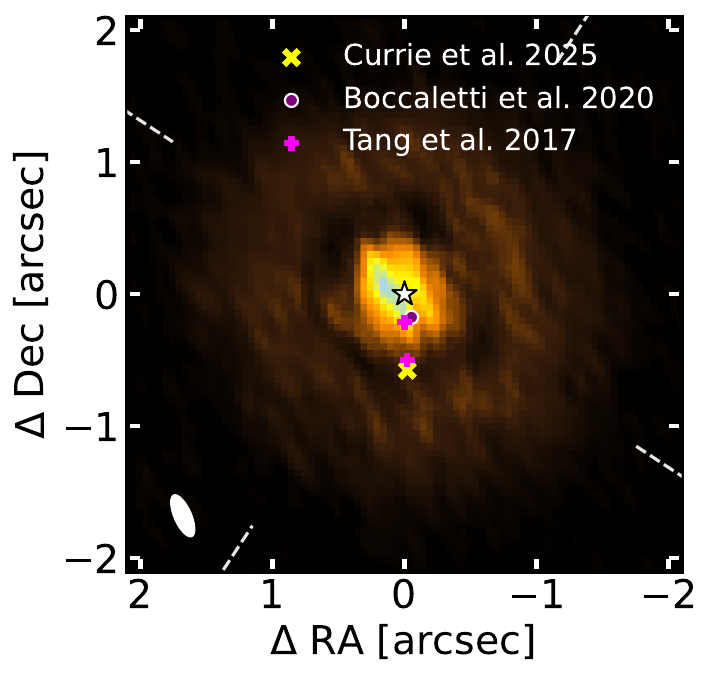}
    \caption{
    The same as \fg{C2H_planet}, but for HCO$^+$ moment 0 map.
    }
    \label{fig:HCOp_planet}
\end{figure*}

\section{Archival molecular data used in this work}\label{app:archival_data}

We summarize here the archival molecular transitions that were published in previous works, and used for rotational diagram analysis in this work. A compilation of the cube information can be found in \tb{transition_archive}, which has a format analogous to to \tb{transition}.

\begin{table*}
\centering
\caption{Molecular data from previous works.}
\label{tab:transition_archive}
\makebox[\textwidth][c]{
\begin{tabular}{l l l l c c c c c c c c}
\hline\hline
Molecule & Transition & Rest freq. & $\log_{10}(A_{ij}/{\rm s^{-1}})$ & $E_u$ & $g_u$ & Beam size (PA) & rms noise & Integrated flux \\
 & & [GHz] & & [K] & & [$''\times''$] ([deg]) & [mJy bm$^{-1}$] &  [mJy km/s] \\

(1) & (2) & (3) & (4) & (5) & (6) & (7) & (8) & (9)\\
\hline
CS$^{a}$    & $J=3-2$                                      & 146.96903  & $-4.21674$ & 14.1 & 7 &  0.79$\times$0.48 (14) & 5.3 & 383$\pm$11
 \\
CS$^{b}$    & $J=7-6$                                      & 342.88285  & $-3.07598$ & 65.8 & 15 &  0.30$\times$0.18 (28) & 1.3 & 370$\pm$75$^{d}$
 \\
SO$^{c}$    & $J_N=6_5-5_4$                                & 219.94944   & $-3.87446$ & 35.0 & 13 & 0.39$\times$0.28 (171) & 1.6 & 623$\pm$12 \\
\hline
\end{tabular}
}
\tablefoot{
Columns are organized in the same way as in \tb{transition}. \\
$^{a}$ from \citet{Rivi`ere-MarichalarEtal2026}. \\
$^{b}$ from \citet{BoothEtal2026}. \\
$^{c}$ from \citet{SpeedieEtal2025}. \\
$^{d}$ The integrated line flux is slightly different from the value reported in \citet[][]{BoothEtal2026}, who gave 440$\pm$44 mJy km/s. The differences are within error range, and due to likely different ways of integration. The integrated line flux of the other two transitions are not reported in the references. For all these three lines, the integrated line flux are measured using the same way as other lines analyzed in this work. The details of measurement was described in \Se{observation}.\\

}
\end{table*}
    
\end{appendix}
\end{document}